\documentstyle[11pt,epsfig]{article}

\newlength{\dinwidth}
\newlength{\dinmargin}
\newcommand{\be}{\begin{equation}}
\newcommand{\bee}{\begin{equation}}
\newcommand{\ee}{\end{equation}}
\newcommand{\beqn}{\begin{eqnarray}}
\newcommand{\eeqn}{\end{eqnarray}}

\newcommand{\rf}{{\bf r}}

\newcommand{\qf}{{\bf q}}

\newcommand{\Gam}{{\Gamma}}
\newcommand{\gam}{{\gamma}}
\newcommand{\lam}{{\lambda}}
\newcounter{savefig}

\def\veps{\mbox{\boldmath$\epsilon$}}
\def\vrho{\mbox{\boldmath$\rho$}}

\begin{document}
\titlepage
\begin{flushright}
CERN-TH/99-51
 \\
MC-TH-99/02 \\
\end{flushright}
\begin{center}
\vspace*{2cm}
{\Large \bf 
Diffractive photon production in $\gamma p$ and $\gamma \gamma$ interactions 
} \\
\vspace*{1cm}

{\Large N.G.Evanson$^1$ and J.R.Forshaw$^2$\footnote{On leave of absence from $^1$.}}

\vspace*{0.5cm}

$^1$Department of Physics and Astronomy,\\ 
University of Manchester,\\
Manchester, M13 9PL. England. \\
\vspace*{0.5cm}
$^2$Theory Division, CERN, \\
1211 Geneva 23. Switzerland.

\end{center}

\vspace*{3cm}
\begin{abstract}
We study the diffractive production of photons in $\gamma p$ and $\gamma
\gamma$ collisions. We specifically compute the rates for $\gam^*p \to
\gam X$ and for $\gam^* \gam^* \to \gam \gam$, where $X$ denotes the proton
dissociation. We focus on the rates at large momentum transfers, 
$-t \gg \Lambda^2$, where we are most confident in the use of QCD 
perturbation theory. However, our calculations do allow us to study the 
$-t \to 0$ behaviour of the $\gamma^* \gamma^* \to \gamma \gamma$ process 
in the region where the incoming photons are sufficiently virtual. 

\vspace*{1.5cm}
PACS numbers: 13.60-r, 13.60.Fz, 13.85-t, 13.85.Dz.
\end{abstract}

\begin{flushleft}
\vfill
CERN-TH/99-51\\
February 1999
\end{flushleft}

\newpage

\section{Introduction}
One of the major challenges in the domain of strong interaction dynamics is
to gain an understanding of diffractive phenomena within the framework of
QCD. Typically, diffractive processes are driven by dynamics at low
momentum transfers and hence cannot be studied using perturbation theory. 

Large-$t$ diffraction has been shown to be one area where perturbation 
theory can be used \cite{FS}. 
Double dissociation in hadron-hadron 
and photon-hadron collisions with a large momentum transfer between the 
diffracting systems has been, and continues to be, studied both theoretically 
\cite{MT,CF} and experimentally \cite{jets}. 
The large momentum transfer means that jets are produced 
in the diffracting systems, the identification of which leads to a sample 
of `gaps between jets' events. A reliable calculation of the expected rate
for double dissociation is made difficult because there is the possibility
of additional strong interactions between the diffracting systems which
could result in the loss of gaps. A process which avoids the `gap survival'
problem is diffractive vector meson production in photon-hadron collisions: 
$\gamma p \to V X$ \cite{Ginzburg,GinzburgIvanov,FRIvanovBFHL,hight}. The meson $V$ 
is produced with a high $p_t$ relative to the incoming photon 
and the largeness of 
this momentum transfer typically leads to the dissociation of the hadron 
into system $X$. A principal challenge in computing the rate for meson 
production lies in determining how the meson is produced. To avoid both 
the meson wavefunction problem and the gap survival problem, one can study 
the diffractive production of photons \cite{Ginzburg,GinzburgIvanov,MWDI}. At HERA, one 
can look for $\gamma p \to \gamma X$ where the
photon is produced with a high $p_t$ and the proton dissociates. At LEP2
or a future linear collider one can look for $\gamma \gamma \to \gamma X$
or $\gamma \gamma \to \gamma \gamma$ \cite{Ginzburg,BRLBrodsky}. In this paper we compute the rates
for diffractive photon production in photon-proton and photon-photon
collisions at large momentum transfer. Our calculations include the 
possibility of off-shell incoming photons.

\section{The $\gamma^{*}\gamma$ impact factor}
The scattering amplitude in the high energy (Regge) limit may be 
factorised in the following way 
\be \label{highenfact}
A(s,t) = \frac{i s \ G}{(2\pi)^4} \int d^2 {\bf k}_1 d^2 {\bf k}_2 
\Phi_1({\bf k}_1,{\bf q}) \Phi_2({\bf k}_2,{\bf q})
f(s,{\bf k}_1,{\bf k}_2,{\bf q}).
\ee

$G$ is the colour factor for the process and the functions 
$\Phi_i({\bf k}_{1},{\bf{q}})$ are the impact factors which contain the 
information about the external hadronic states. The function 
$f(s,{\bf k}_{1},{\bf k}_{2},\bf{q})$ contains all the internal dynamics 
of the BFKL pomeron, is process independent and is well known in the
leading logarithm approximation (LLA) which we use throughout 
\cite{bfkl,BFKLsolve}. The corresponding differential cross-section is
\be
\frac{d\sigma}{dt}= \frac{|A(s,t)|^2}{16\pi s^2}.
\ee

\begin{figure}[h]

\begin{center}
\leavevmode
\hbox{\epsfxsize= 8 cm
\epsfysize= 5 cm
\epsfbox{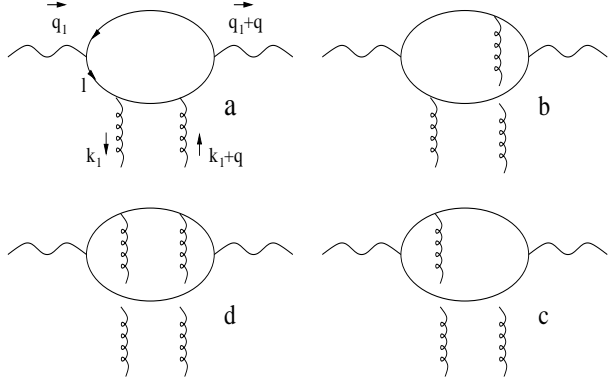}}
\caption{The four diagrams that contribute to the photon-photon impact factor}
\label{fig:fig1}
\end{center}
\end{figure}

For diffractive photon production we need to compute the $\gam^*$-$\gam$ 
impact factor. The relevant Feynman diagrams are shown
in Fig. \ref{fig:fig1}. Dispersive techniques are used to calculate the 
imaginary part of the amplitude which is all that is required in the LLA.
The study of the the photon impact factor is not new and dates back to
\cite{LipQED}. In this paper, we compute the impact factor with an off-shell
incoming photon and a general momentum transfer.

We detail the calculation of the contribution to the impact factor arising 
from the amplitude of Fig.\ref{fig:fig1}(a). The same techniques apply to 
all diagrams. 
\be\label{amp1}
(A^{jk}_{(a)})^{\mu \nu} = -(4\pi)^2\alpha_s\alpha\sum_{q} e^{2}_{q}
\int d(PS) \ {\frac{{\rm Tr}(\not{l}\gamma^{\mu}
(\not{l}-\not{k_{1}})\gamma^{\nu}(\not{l}+\not\!{q})
\not{\epsilon}_{(f)}^k (\not{l}-\not\!{q_{1}})
\not{\epsilon}_{(i)}^j )}{l^{2}(l+q)^{2}}}. 
\ee
The photon polarisation states are labelled by $j$ and $k$ and $d(PS)$ refers
to the two-body phase space of the cut quark lines shown in 
Fig.\ref{fig:fig1}. We define two light-like and mutually orthogonal
vectors, $p_1$ and $p_2$. All momenta are then Sudakov decomposed into 
components proportional to  $p_1$, and $p_2$ and a third perpendicular 
vector:  
\beqn
q_{1}^{\mu} & = & p^{\mu}_{1} - \frac{Q^{2}}{s} p^{\mu}_{2} \nonumber\\
l^{\mu} & = & \rho p^{\mu}_{1}+ \lambda p^{\mu}_{2}+ l^{\mu}_{\perp} \nonumber\\
k^{\mu}_{1} & = & \rho_{1} p^{\mu}_{1}+ \lambda_{1} p^{\mu}_{2} +k_{1\perp}^{\mu} 
\nonumber\\
q^{\mu} & = & \rho_{t} p^{\mu}_{1}+ \lambda_{t} p^{\mu}_{2}+ q^{\mu}_{\perp}.
\eeqn
$q_{1}$ is the incoming photon momentum ($Q^{2}=-q_{1}^{2}$), $l$ the 
quark momentum, $k_{1}+q$ and $-k_{1}$ are the momenta of the gluon legs.
The momentum transfer is $q^2 = t < 0$. Working in the eikonal approximation 
we are only interested in terms 
that are proportional to $p_{1}^{\mu}p_1^{\nu}$ in the numerator of 
(\ref{amp1}). Projecting out this piece, the amplitude becomes 
\be
(A^{jk}_{(a)})^{\mu \nu} = -(4\pi)^2 \frac{4}{s^{2}}\alpha_s\alpha\sum_{q} 
e^{2}_{q} \int d(PS) \ \frac{{\rm Tr}(\not{l}
\not{p_{2}}(\not{l}-\not{k_{1}})\not{p_{2}}(\not\!{l}+
\not\!{q})\not\epsilon_{(f)}^{k} (\not\!{l}-\not\!{q_{1}})
\not\epsilon^{j}_{(i)})}{l^{2}(l+q)^{2}} \ p_1^{\mu} p_1
^{\nu}.
\label{zebedee}
\ee
The two-body phase space integration over the cut lines is 
\beqn
\int d(PS) &=& \int\;\frac{d^{4}l}{(2\pi)^{3}}\;
\frac{d^{4}k_{1}}{(2\pi)^{3}}\;
\delta((l-k_{1})^{2})\;\delta((q_{1}-l)^{2})\nonumber\\
&=&\left(\frac{s}{2}\right)^{2}\!\int d\rho \ d\lambda \ d\rho_{1} \
d\lambda_{1} \;\int \frac{d^{2}{\bf{l}}}{(2\pi)^{3}}
\frac{d^2{\bf k}_1}{(2\pi)^{3}}\;
\frac{-1}{\rho(1-\rho)s^{2}}\;\delta\left(\lambda-\lambda_{1}-
\frac{({\bf{l}}-{\bf k}_{1})^{2}}{s{\rho}}\right)\nonumber\\
&
\times &\delta\left(\lambda+\frac{Q^{2}}{s}+\frac{{\bf{l}}^{2}}{s(1-\rho)}
\right).  
\eeqn
We have used the fact that $\rho_1 \ll \rho$ in the LLA and have transfered 
from the Minkowski metric to the Euclidean one:
\be
-(k^{\mu}_{1\perp})^2= {\bf k}_1^2.
\ee
The $\delta$ functions are used to absorb the $\lam$ and $\lam_1$ integrals
and lead to
\begin{eqnarray}   
q_{1}^{\mu} & = & p^{\mu}_{1} -  Q^{2} \frac{p^{\mu}_{2}}{s}\nonumber\\
l^{\mu} & = & \rho p^{\mu}_{1}- \left(\frac{{\bf{l}}^{2}}{(1-\rho)} + 
Q^{2}\right)\frac{p^{\mu}_{2}}{s}+ l^{\mu}_{\perp} \nonumber\\
k^{\mu}_{1} & = &-\left(\frac{({\bf{k}}_{1}-{\bf{l}})^{2}}{\rho} + 
Q^{2}+\frac{{\bf{l}}^{2}}{(1-\rho)}\right)\frac{ p^{\mu}_{2}}{s}+ 
k^{\mu}_{1\perp} \nonumber\\
q^{\mu} & = & (Q^{2}+{\bf{q}}^{2})\frac{p^{\mu}_{2}}{s}+q_{\perp}^{\mu}.
\end{eqnarray}
Within the LLA it is safe to neglect $\rho_1$ and $\rho_t$.

Our notation follows that of \cite{JRFDR}, and so we can readily
extract the corresponding impact factor by removing the
$$ \frac{2 p_1^{\mu} p_1^{\nu}}{(2 \pi)^4} \int d\rho_1 d^2 {\bf k}_1 $$
from (\ref{zebedee}), i.e. 
\begin{equation}
\Phi_{(a)}^{jk}({\bf k}_{1},{\bf{q}})=-\frac{2\alpha\alpha_{s}}{s^2}
\sum_{q} e^{2}_{q}\int\!\!d^{2}{\bf{l}}\int_{0}^{1}\!\!
d\rho\frac{T_{(a)}^{jk}}{\rho(1-\rho)}
\end{equation}
where $T_{(a)}^{jk}$ is given by
\begin{equation}
T_{(a)}^{jk}=\frac{{\rm Tr}(\not{l}\not{p_{2}}(\not{l}-
\not{k_{1}})\not{p_{2}}(\not\!{l}+\not\!{q}) \not\epsilon_{(f)}^{k}
(\not\!{l}-\not\!{q_{1}}) \not\epsilon^{j}_{(i)})}{l^{2}
(l+q)^{2}}.
\end{equation}

\begin{figure}[h]
\begin{center}
\leavevmode
\hbox{\epsfxsize= 4cm
\epsfysize= 3 cm
\epsfig{file=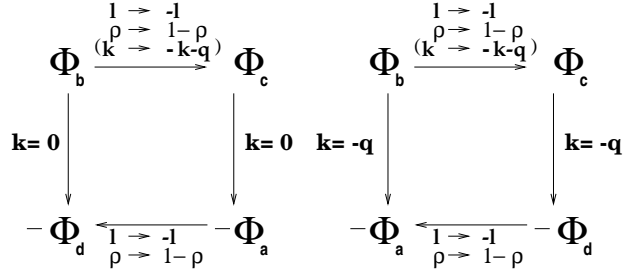}}
\caption{The relationship between the four amplitudes of Fig.\ref{fig:fig1}}
\label{fig:fig2}
\end{center}
\end{figure}

It is not necessary to calculate the contributions to the impact factor 
from all four diagrams explicitly since they are related to each other
by the transformations shown in Fig.\ref{fig:fig2}.

Thus, the full impact factor is given by
\begin{equation}
\Phi_{\gamma \gamma}^{jk}({\bf k}_{1},{\bf{q}})=-
\alpha\alpha_{s}\left(\frac{2}{s} \right)^{2}
\sum_{q} e^{2}_{q}\int\!\!d^{2}{\bf{l}}\int_{0}^{1}\!\!
d\rho\frac{1}{\rho(1-\rho)}
\left[T_{(c)}^{jk}-T_{(c)}^{jk}|_{{{\bf k}_{1}=0}}\right]
\end{equation}
where
\begin{equation}
T_{(c)}^{jk}=\frac{{\rm Tr}(\not{l}\not{p_{2}}(\not{l}+
\not{k_{1}}+\not\!{q})\not\epsilon_{(f)}^{k}(\not\!{l}-\not\!{q_1}+\not{k_1})
\not{p_{2}}(\not\!{l}-\not\!{q_1}) \not\epsilon_{(i)}^{j})}
{(l-q_1)^{2}(l+k_1+q)^{2}}.
\end{equation}

\subsection{Photon polarisation}
The transverse polarisation vectors of the incoming photon
lie in the transverse plane by definition. However, as the final state 
photon gets a kick into the transverse plane, its polarisation vectors have 
a component in the non-transverse directions. In the Lorentz gauge, the 
most general polarisation vectors for the incoming and outgoing photons are

\beqn\label{pol1} 
{\epsilon^k}^\mu_{(f)} \;&=&\;{\epsilon^k_\perp}^\mu\;-
\;\frac{2\;q_\perp \cdot \epsilon^k_\perp}{s}\;p^\mu_2,\nonumber\\
{\epsilon^j}^\mu_{(i)} \;&=&\;{\epsilon^j_\perp}^\mu.
\eeqn
Placing these polarisation vectors into the amplitude and taking the 
trace \cite{FORM} we get   
\beqn\label{impactmom}
\Phi^{jk}_{(x)}({\bf k}_1,{\bf q}) &=& \pm\; 4\alpha_s \alpha
\sum_{q} e_{q}^{2}\int_0^1  \!\!\!d\rho 
\int \! d^2 {\bf l} \, \frac{1}{({\bf l}^2+\overline{Q}^2)}\nonumber\\
&&\hspace{-2.5cm}\frac{ {\bf \Delta}_{(x)} \cdot \veps^{*k}\,
{\bf l}\cdot \veps^j (1-2\rho)^2 - \veps^{*k} 
\cdot \veps^j \, 
{\bf l} \cdot( {\bf l} - {\bf \Delta}_{(x)} )+ 4 {\bf l} \cdot
\veps^{*k} \, {\bf l} \cdot \veps^j \rho(1-\rho) - {\bf l} \cdot 
\veps^{*k} \, {\bf \Delta}_{(x)} \cdot \veps^j} {({\bf \Delta}_{(x)} - 
{\bf l})^2}.
\eeqn
$x=a,b,c,d$ labels the diagram in Fig.\ref{fig:fig1}. The impact factor takes
the $+$ sign for diagrams $(b)$ and $(c)$ and the $-$ sign for diagrams
$(a)$ and $(d)$. We are focusing on diagram $(c)$, in which case
$${\bf \Delta}_{(c)} = {\bf k}_1+\rho {\bf q}.$$
All boldface vectors are Euclidean and lie in the transverse plane, e.g.
${\veps^j}^2 = - {\epsilon^j_\perp}^2$.
We have also introduced $\overline{Q}^{2} \equiv Q^{2}\rho(1-\rho)$.
 
The longitudinal polarisation vector of the incoming photon is given by
\be
{\epsilon^{0}}^{\mu}= {\frac{1}{\sqrt{Q^{2}}}}p_{1}^{\mu} +
\frac{ \sqrt{Q^{2}}}{s}p_{2}^{\mu}.
\ee
After substituting this expression into the amplitude and taking the trace:
\beqn
\Phi^{0i}_{(x)}({\bf k}_{1},{\bf{q}})=\frac{\pm 4\alpha_{s}\alpha}{Q}
\sum_{q} e_{q}^{2} \int_{0}^{1} d\rho \ (2\rho-1)  \int
d^{2}{\bf{l}} \frac{({\bf{l}}^{2}-{\overline{Q}}^{2})}{({\bf{l}}^{2}+
{\overline{Q}}^{2})} \frac{({{\bf \Delta}_{(x)}-\bf{l}})\cdot \veps_i}
{({\bf \Delta}_{(x)}-{\bf{l}})^{2}}
\eeqn
which is identically zero. Thus, in the LLA, longitudinally polarised photons
do not scatter into real photons. 

\subsection{Impact factor in position space}
The four-gluon amplitude, $f(s,{\bf k}_{1},{\bf k}_{2},\bf{q})$ of 
(\ref{highenfact}), can be written \cite{BFKLsolve}:
\beqn
&& f(s,{\bf k}_1,{\bf k}_2,{\bf q}) = 
\frac{1}{(2\pi)^{6}}\;\int\;d^{2} \vrho_1
\;d^{2} \vrho_2\;d^{2} \vrho_3\;d^{2} \vrho_4
\;e^{-i({\bf k}_{1}\cdot \vrho_1 - ({\bf k}_{1} + {\bf{q}})\cdot
\vrho_2-{\bf k}_{2}\cdot \vrho_3 + ({\bf k}_{2} + {\bf{q}})
\cdot \vrho_4)}\nonumber\\
&&\hspace{2.5cm}\int_{-\infty}^{\infty}\;d\nu\; 
\frac{\nu^2}{(\nu^{2}+1/4)^{2}}e^{z\chi(\nu)} E^{\nu}(\vrho_1,\vrho_2)
E^{*\nu}(\vrho_3,\vrho_4).
\eeqn
We have neglected the contributions with conformal spin $n>0$ since they
give rise to sub-leading contributions at high energies. The eigenfunctions 
of the kernel are given explicitly by
\be\label{eigen}
E^\nu(\vrho_{1},\vrho_{2})\;=\;
\left|\frac{\vrho_{1}-\vrho_{2}}{\vrho_{1}\vrho_{2}}\right|^{1+2 i \nu},
\ee
\be
\chi(\nu)=2 \Psi(1)-\Psi(1/2+i\nu)-\Psi(1/2-i\nu)
\ee
($\Psi(x)$ is the digamma function)
and 
\be
z=\frac{N_c \alpha_{s}}{\pi}\ln\left(\frac{s}{s_0}\right).
\ee
The choice of scale $s_0$ is arbitrary in the LLA, in our subsequent 
calculations we choose either the 
largest scale (apart from $s$) in the
problem or set a value for $z$. The number of colours, $N_c = 3$. 

The scattering amplitude (\ref{highenfact}) can now be written
\be\label{ampii}
A(s,t) = {\frac{is \ G}{(2\pi)^{10}}}\int_{-\infty}^{\infty}\;d\nu 
\frac{\nu^2}{(\nu^{2}+1/4)^{2}}e^{z\chi(\nu)}\;I_{1}^{\nu}\;
I_{2}^{*\nu}
\ee
and we have introduced the functions
\be
I_{i}^{\nu}= \int d^2 {\bf k}_{1}\;\Phi_{i}({\bf k}_{1},{\bf{q}})
\int d^2 \vrho_1  d^2 \vrho_2 \ E^\nu(\vrho_{1},\vrho_{2})
\ e^{-i {\bf k}_1 \cdot (\vrho_1-\vrho_2) + i {\bf q} \cdot \vrho_2}.
\label{impact9}
\ee
For photon production we need to calculate (\ref{impact9}) using
$\Phi_{\gamma \gamma}^{jk}$. We work in the helicity basis: 
\be\label{pol2}
\veps^{\pm} = -\frac{1}{\sqrt{2}} (1,\pm i).
\ee
There are four amplitudes to consider:  $(j,k)=(+,+),(+,-),(-,+),(-,-)$
and
\be
{\Phi_{(x)}^{(-,-)}}({\bf k}_{1},{\bf{q}})= 
4\alpha\alpha_{s}\sum_{q}e_{q}^{2}\;\int_{0}^{1} d\rho\
(\rho^2+(1-\rho)^2) \int d^{2}{\bf{l}}\ \frac{1}{({\bf l}^2+
{\overline{Q}}^{
2})} 
\frac{\Delta_{(x)}^{*}l-{\bf l}^{2}}{({\bf\Delta}_{(x)}-{\bf{l}})^2}
\ee
From now on, non-boldface momenta are understood to be complex numbers, e.g.
$l=l_x+il_y$. To compute $I_{(-,-)}^{\nu}$ we first take the Fourier
transform of the impact factor: 
\be
{\Phi_{(x)}^{(-,-)}}(\vrho_1,\vrho_2)=\mbox{e}^{i{\bf{q}}\cdot\vrho_2}\ 
\int d^{2} {\bf k}_{1} \ {\Phi_{(x)}^{(-,-)}}({\bf k}_{1},{\bf{q}})
\;\mbox{e}^{-i\bf{k_1}\cdot(\vrho_1-\vrho_2)}.
\ee
Summing over all four diagrams gives
\beqn
\Phi_{(-,-)}(\vrho_1,\vrho_2)&=& 4\alpha\alpha_{s}\sum_{q}
e_{q}^{2} \ \int_{0}^{1} d\rho \ (\rho^2+(1-\rho)^2) \int 
\mbox{d}^{2}{\bf k}_{1}\ \mbox{d}^{2}{\bf k}_2\
\mbox{e}^{-i{\bf k}_{1}\cdot\vrho_1}\ \mbox{e}^{i{\bf k}_{2}\cdot\vrho_2}
\nonumber\\ &&\hspace{-1cm}\delta^{2}({\bf k}_{2}-{\bf k}_{1}-
{\bf{q}})\int\;\mbox{d}^{2}{\bf{l}}\;\mbox{d}^{2}{\bf{l'}}\
\frac{l'^{*}\ l}{({\bf l}^2+{\overline{Q}}^2) {\bf l}'^2}
\left[\delta^{2}({\bf{l'}}+{\bf{l}}-{\bf k}_{1}-\rho{\bf{q}})\right.
\nonumber\\ &&\left.\hspace
{-1.5cm}-\delta^{2}({\bf{l'}}-{\bf{l}}-
{\bf k}_{1}-(1-\rho){\bf{q}}) - \delta^{2}({\bf{l'}}+{\bf{l}}-
\rho{\bf{q}})+\delta^{2}({\bf{l'}}-{\bf{l}}-(\mbox{1}-\rho){\bf{q}})
\right].
\eeqn
Equivalently, and dropping for the time being the pre-factors and the
integral over the momentum fraction $\rho$, we have
\beqn
&&\hspace{-1.5cm}\int\;\frac{d^{2}{\bf r}_{1}\ d^{2}{\bf r}_{2}}{(2\pi)^4} \
\int\;\mbox{d}^{2}{\bf k}_{1}\ \mbox{d}^{2}{\bf k}_{2} \ \mbox{e}
^{-i\bf{k_1}\cdot\vrho_1}\;\mbox{e}^{i\bf{k_2}\cdot\vrho_2}\
\mbox{e}^{i{\bf r}_{1}\cdot( \bf{k_2-\bf{k_1}}-\bf{q})}\
\int\ d^{2}{\bf{l}}\ d^{2}{\bf{l'}}\ \frac{l'^{*}\;l}{({\bf l}^2+
\overline{Q}^2)\ {\bf l}'^2}\nonumber\\
&&\left[\mbox{e}^{-i{\bf r}_{2}\cdot({\bf{l'}}+{\bf{l}}-
{\bf k}_{1}-\rho{\bf{q}})}  -  \mbox{e}^{i{\bf r}_{2}\cdot({\bf{l'}}-
{\bf{l}}-{\bf k}_{1}-(1-\rho){\bf{q}})} - \mbox{e}^{i{\bf r}_{2}
\cdot({\bf{l'}}+{\bf{l}}-\rho{\bf{q}})} + \mbox{e}^{i{\bf r}_{2}
\cdot({\bf{l'}}-{\bf{l}}-(\mbox{1}-\rho){\bf{q}})}\right].
\eeqn
After some algebra, 
this becomes
\beqn
&&\epsilon_{a}^{-}\epsilon_{b}^{+}\ \int\ \frac{d^{2}{\bf r}_{1}\
d^{2}{\bf r}_{2}}{(2\pi)^4}\ \mbox{e}^{-i{\bf r}_{1}\cdot{\bf{q}}}\;\int\;
\mbox{d}^{2}{\bf k}_{1}\;\mbox{d}^{2}{\bf k}_{2}\left[\;\mbox{e}^{i
\bf{k_1}\cdot({\bf r}_{1}+{\bf r}_{2}+\vrho_1)}-\;
\mbox{e}^{-i\bf{k_1}\cdot({\bf r}_{1}+\vrho_1)}\right]\;\mbox{e}^
{i\bf{k_2}\cdot({\bf r}_{1}+\vrho_2)}\nonumber\\
&&\hspace{-0.8cm}\left[\mbox{e}^{-i{\bf r}_{2}
\cdot\rho{\bf{q}}}\left(\frac{\nabla_{{\bf r}_{2}}^{a}}{i}\int\;d^{2}
{\bf{l}}\;\frac{\mbox{e}^{i{\bf r}_{2}\cdot{\bf{l}}}}{{\bf l}^2+
\overline{Q}^2}\right)-\mbox{e}^{-i{\bf r}_{2}\cdot(1-\rho){\bf{q}}}\left
(\frac{\nabla_{{\bf r}_{2}}^{a}}{-i}\int\;d^{2}{\bf{l}}\;
\frac{\mbox{e}^{-i{\bf r}_{2}\cdot{\bf{l}}}}{{\bf l}^2+\overline{Q}^2}
\right)\right]\;\frac{\nabla_{{\bf r}_{2}}^{b}}{i}\int\;d^{2}{\bf{l'}}\;
\frac{\mbox{e}^{-i{\bf r}_{2}\cdot{\bf{l'}}}}{{\bf l}'^2}
\eeqn
where $a$ and $b$ are vector indices in the transverse space.
Performing the angular parts of the $\bf{l}$ and  $\bf{l'}$ integrals and 
the ${\bf k}_1$ and ${\bf k}_2$ integrals yields
\beqn
&&-\epsilon_{a}^{-}\epsilon_{b}^{+}\ (2\pi)^{2}\int\ 
d^{2}{\bf r}_{1}\ d^{2}{\bf r}_{2}\left [
\mbox{e}^{-i{\bf r}_{2}\cdot\rho{\bf{q}}}+\;\mbox{e}^{-i{\bf r}_{2}
\cdot(1-\rho){\bf{q}}}\right]\;\mbox{e}^{-i{\bf r}_{1}\cdot{\bf{q}}}
\left(\nabla_{{\bf r}_{2}}^{a}\int\;d|{\bf l}|\ \frac{|{\bf l}|
J_0(|{\bf r}_2| |{\bf l}|)} {{\bf l}^2+\overline{Q}^2}\right)\nonumber\\
&&\left(\nabla_{{\bf r}_{2}}^{b}\int\;d|{\bf l}'| \
\frac{J_0(|{\bf r}_{2}||{\bf l}'|)}{|{\bf l}'|}\right)\left[\delta^{2}
({\bf r}_{1}+{\bf r}_{2}+\vrho_1)-\delta^{2}({\bf r}_{1}+\vrho_1)\right]\
\delta^{2}({\bf r}_{1}+\vrho_2).
\eeqn
Using
\be
\nabla_{{\bf r}_{2}}J_0(|{\bf r}_{2}||{\bf l}|)=- |{\bf l}|\ 
J_1(|{\bf r}_{2}| |{\bf l}|)\ \hat{{\bf r}}_{2},
\ee
\be
\veps^{+}\cdot\hat{{\bf r}}_{2}\ \veps^{-}\cdot\hat{{\bf r}}_{2}=1,
\ee
completing the radial ${\bf l}$ integral and re-arranging leads to the
result:
\beqn
\Phi_{(-,-)}(\vrho_1,\vrho_2)&=&
16\pi^{2}\alpha\alpha_s\ \sum_{q}e_{q}^{2}
\int_0^1
 d \rho\ [\rho^2+(1-\rho)^2]\ \int d^2 \rf_1 d^2 \rf_2 \ 
\frac{1}{|\rf|}\ \mbox{e}^{i \rho \qf \cdot \rf_1}\;\mbox{e}^{i (1-\rho) \qf 
\cdot \rf_2}\;\nonumber\\ &&\ {\overline{Q}}  K_{1}(|\rf|{\overline{Q}}) \
[\delta^2(\rf_1-\vrho_1)-\delta^2(\rf_2-\vrho_1)]\
[\delta^2(\rf_1-\vrho_2)-\delta^2(\rf_2-\vrho_2)],
\eeqn
where ${\bf r} = {\bf r}_1 - {\bf r}_2$.
The impact factors for the other polarisation states are  
\beqn\label{impact}
\Phi_{(+,+)}(\vrho_1,\vrho_2)&=&\Phi_{(-,-)}(\vrho_1,\vrho_2)\nonumber\\
\Phi_{(+,-)}(\vrho_1,\vrho_2)&=&32\pi^{2}\alpha\alpha_s\ \sum_{q}
e_{q}^{2} \int_0^1 d \rho\ \int d^2 \rf_1 d^2 \rf_2\ \rho(1-\rho)\
\frac{r^2}{|\rf|^3}\ \mbox{e}^{i \rho \qf \cdot \rf_1}\ \mbox{e}^{i (1-\rho) 
\qf \cdot \rf_2}\ \nonumber\\ && K_{1}(|\rf|{\overline{Q}})\ {\overline{Q}}\
[\delta^2(\rf_1-\vrho_1)-\delta^2(\rf_2-\vrho_1)]\
[\delta^2(\rf_1-\vrho_2)-\delta^2(\rf_2-\vrho_2)]\nonumber \\
\Phi_{(-,+)}(\vrho_1,\vrho_2)&=&32\pi^{2}\alpha\alpha_s\ 
\sum_{q}e_{q}^{2} \int_0^1 d \rho
\ \int d^2 \rf_1 d^2 \rf_2\
\rho(1-\rho)\ \frac{{r^*}^2}{|\rf|^3}\ \mbox{e}^{i \rho \qf \cdot \rf_1}\
\mbox{e}^{i (1-\rho) \qf \cdot \rf_2}\ \nonumber\\ &&
K_{1}(|\rf|{\overline{Q}})\ {\overline{Q}}\
[\delta^2(\rf_1-\vrho_1)-\delta^2(\rf_2-\vrho_1)]\
[\delta^2(\rf_1-\vrho_2)-\delta^2(\rf_2-\vrho_2)].
\eeqn

\section{Projection on Conformal Eigenstates}
We must now project the position space impact factors onto the conformal 
eigenfunctions of the BFKL kernel, i.e. we must perform the following 
convolution:
  
\be
I^{\nu}_{(h^*_\gamma,h_\gamma)}(Q^{2},q^{2})=\int d^2 \vrho_1  d^2 \vrho_2\ 
\Phi_{(h^*_\gamma,h_\gamma)}(\vrho_1,\vrho_2)\ E^\nu(\vrho_{1},\vrho_{2}).
\ee
This part of the calculation follows very closely that of \cite{MWDI}. 
For completeness, we reproduce the steps here.

In two of the terms of (\ref{impact}) the $\delta$-functions force 
$\vrho_1=\vrho_2$ which results in a vanishing contribution due to the
vanishing of $E^\nu$. Hence,
\beqn 
&&I^{\nu}_{(+,+)}(Q^{2},q^{2})= -32\pi^{2}\alpha_s \alpha 
\sum_{q}e_{q}^{2} \int_0^1 d \rho\ [\rho^2+(1-\rho)^2]\
{\overline{Q}}\nonumber\\ && \int d^2 \rf_1 d^2 \rf_2 \
\frac{1}{|\rf|}\ K_{1}(|\rf|{\overline{Q}}) \
{e}^{i \rho \qf \cdot \rf_1}\ {e}^{i (1-\rho) \qf \cdot \rf_2}\
\left|\frac{\rf}{\rf_1 \rf_2}\right|^{1+2 i \nu}
\eeqn
i.e.
\beqn\label{rho_0} 
&&I^{\nu}_{(+,+)}(Q^{2},q^{2})=-32\pi^{2}\alpha_{s}\alpha
\sum_{q}e_{q}^{2} \ \int_0^1 d \rho\
[\rho^2+(1-\rho)^2]\ {\overline{Q}}\nonumber\\
&&\ \int d^2 \rf\ d^2 \rf_2\
\mbox{e}^{i \rho
 \qf \cdot \rf}\;\mbox{e}^{i \qf \cdot \rf_2}
|\rf|^{2i\nu}\ K_{1}(|\rf|{\overline{Q}})\
|(\rf+\rf_2)\ \rf_2|^{-1-2 i \nu}\ .
\eeqn
We next shift $\rf_2$ by $-\rho \rf$ to eliminate the $\rf$ dependent phase 
factor:
\beqn
&&I^{\nu}_{(+,+)}(Q^{2},q^{2})=-32\pi^{2}\alpha_{s}\alpha
\sum_{q}e_{q}^{2}\int_0^1 d \rho\ [\rho^2+(1-\rho)^2]\
{\overline{Q}}\ \int d^2 \rf\ d^2 \rf_2\ \mbox{e}^{i \qf \cdot \rf_2}
\nonumber\\ &&\hspace{2cm}K_{1}(|\rf|{\overline{Q}}) \ |\rf|^{2i\nu}\
|[\rf_2+(1-\rho)\rf]\ [\rf
_2-\rho\rf]|^{-1-2 i \nu}.
\eeqn
We now switch to the complex notation, i.e from
$\rf=(r_1,r_2)$ to $a=r_1+i\;r_2$ and $b=r_1-i \;r_2$,
and use the freedom to choose $q$ to be real:
\beqn
&&I^{\nu}_{(+,+)}(Q^{2},q^{2})= -8\pi^{2}\alpha_{s}\alpha
\sum_{q}e_{q}^{2} \int_0^1 d \rho\ [\rho^2+(1-\rho)^2]\
{\overline{Q}}\ \int da \ db\ da_2\ db_2 \ \mbox{e}^{i q/2 (a_2+b_2)}
\nonumber\\ && (a b)^{i\nu}K_{1}(\sqrt{a b\ {\overline{Q}}^{2}})\ \{ 
[a_2+(1-\rho)a]\ [b_2+(1-\rho)b]\ [a_2-\rho a][b_2-\rho b]\}^{-1/2- i \nu}.
\eeqn
To proceed we factorise the integration in $a,b$ and $a_2,b_2$ by writing 
the Bessel function as its Mellin transform,
\begin{equation}
K_{1}(x)=\frac{1}{8\pi}\int_{-\infty}^{\infty} d \lambda\ 
\left(\frac{x}{2}\right)^{-\gam-i\lam} \Gam(\frac{i\lam+\gam-1}{2})\
\Gam(\frac{i\lam+\gam+1}{2}) \hspace*{1cm}  [\Re e \ \gam > 1],
\end{equation}
and re-scaling $a$ \& $b$ by $a_2$ \& $b_2$ respectively:
\beqn 
&&I^{\nu}_{(+,+)}(Q^{2},q^{2})=-2\pi\alpha_{s}\alpha\ \sum_{q}
e_{q}^{2}\int_0^1 d \rho\ [\rho^2+(1-\rho)^2]\ \int_{-\infty}^{\infty} 
d\lambda\ ({\overline{Q}}^{2})^{\frac{-\gam-i\lam+1}{2}}\nonumber\\
&&2^{\gam+i\lam-1} \Gam\left(\frac{i\lam+\gam-1}{2} \right)\ 
\Gam\left(\frac{i\lam+\gam+1}{2}\right) \ 
\int da_2 \ db_2 \ \mbox{e}^{i q/2 (a_2+b_2)}\ (a_2 b_2)^{\frac{-\gam-
i\lam}{2}-i\nu} \\ &&\hspace{0cm}\int d a \ d b\ (a b)^{\frac{-\gam-i\lam}{2}
+i\nu}\ \{ [1+(1-\rho)a]\ [1+(1-\rho)b]\ [1-\rho a][1-\rho b]\}^{-1/2- i \nu}.
\nonumber
\eeqn
Re-scaling $a_2$ and $b_2$ by 
$2 i/q$ gives
\beqn \label{rescaled}
&&I^{\nu}_{(+,+)}(Q^{2},q^{2})=2\pi\alpha_{s}\alpha\ \sum_{q}
e_{q}^{2}\left(\frac{2}{q}\right)^{1-2i\nu}\ \int_0^1 d \rho\
[\rho^2+(1-\rho)^2]\ \int_{-\infty}^{\infty} d \lambda\ \left(
\frac{{\overline{Q}}^{2}}{q^2}\right)^{\frac{-\gam-i\lam+1}{2}}\nonumber\\
&&\hspace{2cm}\Gam\left(\frac{i\lam+\gam-1}{2}\right)\ 
\Gam\left(\frac{i\lam+\gam+1}{2}\right) \
\int da_2 \ db_2\ \mbox{e}^{-(a_2+b_2)}\ (-a_2 b_2)^{\frac{-\gam-i\lam}{2}-
i\nu}\nonumber \\ &&\hspace{1
cm}\int da \ db \ (a b)^{\frac{-\gam-i\lam}{2}+
i\nu} \ \{ [1+(1-\rho)a]\ [1+(1-\rho)b]\ [1-\rho a][1-\rho b]\}^{-1/2- i \nu}.
\eeqn
Completing the $a,b$ and $a_2,b_2$ integrals as in \cite{MWDI}:
\beqn\label{ab1}
&&\int d a_2 \ d b_2\ 
\mbox{e}^{-(a_2+b_2)} (-a_2\;b_2)^{\frac{-\gam-i\lam}{2}-i\nu}\nonumber\\
&=& 2 i \ \sin[(\frac{\gam+i\lam}{2}+i\nu)\pi] 
\int_0^{\infty} da_2 \ a_2^{\frac{-\gam-i\lam}{2}-i\nu}\ \mbox{e}^{-a_2}
\int_0^{\infty} db_2 \ b_2^{\frac{-\gam-i\lam}{2}-i\nu}\ \mbox{e}^{-b_2}
\\ &=&2
 i\ \sin[(\frac{\gam+i\lam}{2}+i\nu)\pi] \
\Gamma^2\left( \frac{-\gam-i\lam}{2}-i\nu+1 \right) \nonumber
\eeqn
\beqn
&&\int da \ db \ (a b)^{\frac{-\gam-i\lam}{2}+i\nu}
\{ [1+(1-\rho)a] [1+(1-\rho)b] [1-\rho a][1-\rho b]\}^{-1/2- i \nu}
\nonumber\\ &&=2 i \sin[(-1/2-i\nu)\pi]
\left\{ \int_{-1/(1-\rho)}^0 db \ \int_{-\infty}^{-1/(1-\rho)} da
+\int_0^{1/\rho} db \ \int^{\infty}_{1/\rho} da \right\}\nonumber\\
&&\hspace{.5cm}(a b)^{\frac{-\gam-i\lam}{2}+i\nu} \{-[1+(1-\rho)a]
[1+(1-\rho)b] [1-\rho a][1-\rho b]
\}^{-1/2- i \nu} \nonumber\\
&&=2i \ \sin(z_2\pi) \ [\rho(1-\rho)]^{z_2} (1-\rho)^{-2(1+z_1+z_2)}
\int_0^1 db \ b^{z_1} \left\{(1-b)\left(1+\frac{\rho}{1-\rho}b\right)
\right\}^{z_2} \nonumber \\
&&\hspace{.5cm}\int_{1}^{\infty} da\ a^{z_1}
\left\{(a-1)\left(a+\frac{1-\rho}{\rho}\right)\right\}^{z_2} +
\left[\rho\rightarrow(1-\rho)\right]
\label{nasty}
\eeqn
and we have defined
\beqn
z_{1}&=&\frac{-\gam-i\lam}{2}+i\nu\nonumber\\
z_{2}&=&-1/2-i\nu.
\eeqn
The above integrals are representations of the hypergeometric function,
i.e. (\ref{nasty}) can be written
\beqn
&=&4i\ \sin(z_2\pi)[\rho(1-\rho)]^{z_2}\frac{\Gam(z_1+1)\Gam(z_2+1)}
{\Gam(z_1+z_2+2)}\frac{\Gam(-z_1-2z_2-1)\Gam(z_2+1)}{\Gam(-z_1-z_2)}
(1-\rho)^{-2(1+z_1+z_2)}\nonumber\\
&&\mbox{$_2$}F_1\left(-z_2,z_1+1;z_1+z_2+2;\frac{\rho}{\rho-1}\right)
\mbox{$_2$}F_1\left(-z_2,-z_1-2z_2-1;-z_1-z_2;\frac{\rho-1}{\rho}\right)
\eeqn
and we have anticipated the symmetry between $\rho$ and $(1-\rho)$ when 
performing the $\rho$ integral. 
We can further simplify the hypergeometric functions to reach
\beqn\label{ab}
&=&-4i\ \sin((1/2+i\nu)\pi)\ \frac{\Gam^2(1/2-i\nu)\,
\Gam(\frac{-\gam-i\lam}{2}+i\nu+1)\,\Gam(\frac{\gam+i\lam}{2}+i\nu)}
{\Gam(\frac{\gam+i\lam+1}{2})\Gam(\frac{-\gam-i\lam+3}{2})}
(1-\rho)^{-1+\gam+i\lam}\nonumber\\ &&\mbox{$_2$}F_1\left(1/2+i\nu,1/2-i\nu;
\frac{-\gam-i\lam+3}{2}; \rho\right)\mbox{$_2$}F_1
\left(1/2+i\nu,1/2-i\nu;\frac{\gam+i\lam+1}{2};1-\rho\right).
\nonumber \\
\eeqn
Using these integrals, (\ref{rescaled}) reduces to
\beqn\label{nonf
lip} 
&&I^{\nu}_{(+,+)}(Q^{2},q^{2})=16\pi^{3}\alpha_{s}\alpha\
\sum_{q}e_{q}^{2}\left(\frac{2}{q}\right)^{1-2i\nu} \
\frac{\Gam(1/2-i\nu)}{\Gam(1/2+i\nu)}\ \int_{-\infty}^{\infty} d \lambda\;
\left(\frac{Q^2}{q^2}\right)^{\frac{-\gam-i\lam+1}{2}}\ 
\frac{\Gam(\frac{i\lam+\gam-1}{2})}{\Gam(\frac{-i\lam-\gam+3}{2})}\nonumber\\
& &\Gam(\frac{-\gam-i\lam}{2}-i\nu+1)\ \Gam(\frac{-\gam-i\lam}{2}+i\nu+1)\
\int_0^1 d \rho\ [\rho^2+(1-\rho)^2]\ \left(\frac{\rho}{1-\rho}
\right)^{\frac{-\gam+1-i\lam}{2}}\nonumber\\ &&\mbox{$_2$}F_1
\left(1/2+i\nu,1/2-i\nu;\frac{-\gam-i\lam+3}{2}; \rho\right)
\mbox{$_2$}F_1\left(1/2+i\nu,1/2-i\nu;\frac{\gam+i\lam+1}{2};1-\rho\right)
\eeqn
and we have used the identity
\be
\Gam(z)\Gam(1-z)=\frac{\pi}{\sin(z\pi)}.
\ee 
The $\rho$-integration is performed by expanding the first hypergeometric 
function, performing the integral and then reducing the remaining series 
using $\Psi$ (digamma) functions. The final result for the non-flip 
amplitude takes the following form:
\beqn\label{nonflip1}
&&I^{\nu}_{(+,+)}(Q^{2},q^{2})=\frac{-i}{2}\alpha_{s}\alpha\pi^{5}\sum_{q}e_{q}^{2}
\left(\frac{2}{q}\right)^{1-2i\nu}\ 
\frac{1}{\Gam^{2}(1/2+i\nu)}\;\frac{\tanh(\pi\nu)}{\pi\nu}\frac{1}{(1+\nu^2)}
\int_{\gam-1-i\infty}^{\gam-1+i\infty} dz\nonumber\\
&&\left(\frac{Q^2}{q^2}\right)^{-z/2}\Gam(1/2-i\nu-z/2)\
\Gam(1/2+i\nu-z/2)\;\Gam(z/2)\ \Gam(z/2+1)\ (z^{2}+11+12\nu^2)
\eeqn
and we have changed integration variables to $z=\gam-1+i\lam.$ Of course
the $(-,-)$ amplitude is equivalent.
For the flip amplitudes we proceed in the same fashion. The final answer
is
\beqn\label{flip1}
I^{\nu}_{(+,-)}(Q^{2},q^{2}) &=& I^{\nu}_{(-,+)}(Q^{2},q^{2}) 
= -2i\alpha_s\alpha\pi^{5}\sum_{q}e_{q}^{2}
\left(\frac{2}{q}\right)^{1-2i\nu}\ \frac{1}{\Gam^{2}(1/2+i\nu)}\
\frac{\tanh(\pi\nu)}{\pi\nu}\frac{1}{(1+\nu^2)} \nonumber \\ & & 
\hspace*{-1.5cm} \int_{\gam-1-i\infty}^{\gam-1+i\infty} dz 
\left(\frac{Q^2}{q^2}\right)^{-z/2}\Gam(3/2-i\nu-z/2)\
\Gam(3/2+i\nu-z/2)\;\Gam(z/2)\ \Gam(z/2+1).
\eeqn

\subsection{The limits $Q^{2}\gg q^{2}$ and $Q^{2}\ll q^{2}$}
In the limit $Q^{2}\gg q^{2}$ we are able to complete the $z$ integration. 
We must close the contour in the right 
half plane. The leading contribution is extracted by considering the left 
most poles.

For the non-flip amplitude these are located at $z=1\pm2i\nu$:
\beqn\label{nonfliplim1}
&&I^{\nu}_{(+,+)}(Q^{2},q^{2}){\Big |}_{Q^{2}\gg q^{2}}=2\alpha_{s}
\alpha\pi^{6}\sum_{q}e_{q}^{2}\left(\frac{2}{q}\right)^{1-2i\nu}
\frac{1}{\Gam^{2}(1/2+i\nu)}\;\frac{\tanh(\pi\nu)}{\pi\nu}\frac{1}{(1+\nu^2)}
\nonumber\\ &&\left\{\left(\frac{Q^2}{q^2}\right)^{-1/2-i\nu}
\Gam(-2i\nu)\Gam(1/2+i\nu)\Gam(3/2+i\nu)[(1+2i\nu)^{2}+11+12\nu^{2}]\; + \; 
c.c. \right\}.
\eeqn
For the flip amplitude, the left most poles are located at $z=3\pm2i\nu$ 
giving
\beqn\label{fliplim1}
&&I^{\nu}_{(+,-)}(Q^{2},q^{2}){\Big |}_{Q^{2}\gg q^{2}}=8\alpha_{s}
\alpha\pi^{6}\sum_{q}e_{q}^{2} \left(\frac{2}{q}\right)^{1-2i\nu}\
\frac{1}{\Gam^{2}(1/2+i\nu)}\;\frac{\tanh(\pi\nu)}{\pi\nu}\frac{1}{(1+\nu^2)}
\nonumber\\ &&\hspace{2cm}\left\{\left(\frac{Q^2}{q^2}\right)^{-3/2-i\nu}
\Gam(-2i\nu)\Gam(3/2+i\nu)\Gam(5/2+i\nu)\;+\; c.c.\right\}.
\eeqn
We see that, in the limit $Q^{2}\gg q^{2}$, the flip amplitude is suppressed 
by a factor of $q^2/Q^2$.

In the limit $Q^{2}\ll q^{2}$ we must close the contour in the left hand 
plane. The leading piece is found by considering the pole at $z=0$. 
This gives the $Q^2$ independent contribution found in \cite{MWDI}. 
The first correction to this is found by considering the double pole located 
at $z=-2$. We find 
\beqn\label{nonfliplim2}
&&I^{\nu}_{(+,+)}(Q^{2},q^{2}){\Big |}_{Q^{2}\ll q^{2}}=2\alpha_{s}
\alpha\pi^{6}\sum_{q}e_{q}^{2} \left(\frac{2}{q}\right)^{1-2i\nu}\
\frac{\Gam(1/2-i\nu)}{\Gam(1/2+i\nu)}\ \frac{\tanh(\pi\nu)}{\pi\nu}\frac{1}
{(1+\nu^2)}\left[(11+12\nu^2)\right.\nonumber\\
&&\left.-(\nu^{2}+1/4)\;\frac{Q^2}{q^2}\ (15+12\nu^{2})\;\left
\{\ln\frac{|t|}{Q^2}-\frac{8}{15+12\nu^{2}}-\Psi(3/2+i\nu)-\Psi(3/2-i\nu)
\right\}\right]. 
\eeqn
The analytic structure of the $z$ plane is identical for the flip result 
and so 
\beqn\label{fliplim2}
&&I^{\nu}_{(+,-)}(Q^{2},q^{2}){\Big |}_{Q^{2}\ll q^{2}}=
8\alpha_{s}\alpha\pi^{6}\sum_{q}e_{q}^{2} 
\left(\frac{2}{q}\right)^{1-2i\nu}\ \frac{\Gam(1/2-i\nu)}{\Gam(1/2+i\nu)}\
\frac{\tanh(\pi\nu)}{\pi\nu}\frac{(\nu^{2}+1/4)}{(1+\nu^2)}\nonumber\\
&&\hspace{1.5cm}\left[1-(\nu^{2}+9/4)\ \frac{Q^2}{q^2}\
\left\{\ln\frac{|t|}{Q^2}-\Psi(5/2+i\nu)-\Psi(5/2-i\nu)\right\}\right
]. 
\eeqn
\section{$\gamma^{*}\gamma^{*}\rightarrow\gamma\gamma$}
The cross section for $\gamma^{*}\gamma^{*}\rightarrow\gamma\gamma$ 
scattering can be computed by substituting the results for $I^{\nu}$ given by  
(\ref{nonflip1}) and (\ref{flip1}) into (\ref{ampii}). The colour factor for 
this process is given by
\begin{equation}
G = \frac{N_{c}^{2}-1}{4} = 2.
\end{equation}

\subsection{The small $|t|$ limit}
For this process we can study the $|t|\rightarrow 0$ region. The cross 
section will be greatest in this limit and, interestingly, we can study the 
behaviour of the amplitude in $|t| \to 0$ limit. We focus on the limit
$Q_{1}^{2} \sim Q_{2}^{2} \gg |t|$ where $Q_1^2$ and $Q_2^2$ are the 
virtualities of the incoming photons. We shall find an analytic approximation 
for the $\gamma^{*}\gamma^{*}\rightarrow\gamma\gamma$ amplitude using the 
approximation to $I^{\nu}$ given by equation (\ref{nonfliplim1}) 
and evaluating the $\nu$ integration using the saddle point 
method. We need only consider the non-flip amplitudes. When 
(\ref{nonfliplim1}) is placed into (\ref{ampii}) we find two distinct terms. 
One term is independent of $|t|$ and gives the limiting value of the 
amplitude when $|t|=0$. The other term provides the leading $|t|$ dependence. 
Considering the saddle point integration of the $|t|$ dependent piece we find 
that the dominant term in the exponential is proportional to 
$\ln(Q_{1}^{2} Q_{2}^{2}/t^2)$ and so we may make the assumption that the 
saddle point lies at $\nu\approx \pm i/2$ and approximate $\chi(\nu)$ by an 
expansion about $\nu=\pm i/2$: $\chi(\nu)\approx 1/(1/2\pm i\nu)$. 
The saddle point is in fact at
\be
\nu=\pm\frac{i}{2}\left(1-\sqrt{\frac{4z}{\ln\left(Q_{1}^{2}Q_{2}^{2}/t^{2}
\right)}}\right)\approx\pm\frac{i}{2}.    
\ee
For the $|t|$ independent piece we have a term proportional to 
$\ln(Q_{1}^{2}/Q_{2}^{2})$ in the exponential. For $Q_1^2 \sim Q_2^2$
the saddle point is at 
\be
\nu=\pm\frac{i\ln(Q_{1}^{2}/Q_{2}^{2})}{28\zeta(3)z}\approx 0
\ee
and we can use $\chi(\nu)
\approx 4\log{2}-14\zeta(3)\nu^{2}$. 
Performing the integration over $\nu$ yields
\beqn\label{Q2ggtgg}
&&A_{(+,+;+,+)}(s,t,Q_{1}^{2},Q_{2}^{2}){\Big |}_{Q_{1}^{2}
\sim Q_{2}^{2}\gg|t|} \approx i \ 9 \pi^{2}\alpha^{2}\alpha_{s}^{2} 
\left(\sum_q e_q^2 \right)^2 \ \frac{s}{Q_{1}Q_{2}}
\left[\exp\left(4z\ln 2-\frac{\ln^{2}\left(Q_{1}^{2}/Q_{2}^{2}\right)}
{56\zeta(3)z}\right)\right.\nonumber\\
&&\hspace{2cm}\left.\sqrt{\frac{\pi}{14\zeta(3)z}}-\frac{4}{9\pi^{7/2}}
\frac{|t|}{Q_{1}Q_{2}}\frac{\ln^{5/4}\left(Q_{1}^{2}Q_{2}^{2}/|t|^{2}\right)}
{z^{7/4}}\exp\left(2\sqrt{z\ln(Q_{1}^{2}Q_{2}^{2}/|t|^{2})}\right)\right].
\eeqn

There is a cusp in the cross-section at $-t=0$, i.e. the slope is
infinite. In particular, defining an effective $b$-parameter 
$b_{\rm eff}$ by
\be
b_{\rm eff} = \frac{2}{A(s,t)}\ \frac{\partial A(s,t)}{\partial t}
\ee
and using (\ref{Q2ggtgg}), we find that, in the vicinity of $-t=0$,
\be
b_{\rm eff} = \frac{1}{A_0} \frac{4}{9 \pi^{7/2}} \frac{1}{Q_1 Q_2} 
\exp\left(2\sqrt{z\ln(Q_{1}^{2}Q_{2}^{2}/|t|^{2})}\right) \
\frac{\ln^{5/4}\left(\frac{Q_1^{2} Q_2^{2}}{-t}\right)}{z^{7/4}}  
\ee
where
\be
A_0 \approx e^{4z\ln 2} \ \sqrt{\frac{\pi}{14\zeta(3)z}}.
\ee

\subsection{The large $|t|$ limit}
We now consider the limit $Q_{1}^{2}, Q_{2}^{2}\ll|t|$. The saddle point is
at $\nu=0$ and, after integration over $\nu$, we find that
\begin{eqnarray}
A_{(+,+;+,+)}(s,t,Q_{1}^{2},Q_{2}^{2}){\Big |}_{|t|\gg Q_{1}^{2},  
Q_{2}^{2}} &=& \frac{i}{4} \pi \frac{s}{|t|} \alpha^2 \alpha_s^2 \
\left(\sum_{q} e_q^2 \right)^2 \
e^{4z\ln 2}\ \left(\frac{\pi}{14\zeta(3)z}\right)^{3/2}
C_1(Q_1) C_1(Q_2) \nonumber \\
A_{(+,-;+,-)}(s,t,Q_{1}^{2},Q_{2}^{2}){\Big |}_{|t|\gg Q_{1}^{2}, 
Q_{2}^{2}} &=& \frac{i}{4} \pi \frac{s}{|t|} \alpha^2 \alpha_s^2 \ 
\left(\sum_{q} e_q^2 \right)^2 \
e^{4z\ln 2}\ \left(\frac{\pi}{14\zeta(3)z}\right)^{3/2} 
C_2(Q_1) C_2(Q_2) \nonumber \\
\end{eqnarray}
where
\beqn
C_1(Q_1) &=&  11+\frac{1}{4}\frac{Q_{1}^{2}}{|t|}
\left(68-30\gam_{E}-60\ln 2-15\ln\left(\frac{|t|}{Q_{1}^{2}}\right)\right)
\nonumber \\ 
C_2(Q_1) &=&  1-\frac{9}{4}\frac{Q_{1}^{2}}{|t|}
\left(-\frac{16}{3}+2\gam_{E}+4\ln 2+\ln\left(\frac{|t|}{Q_{1}^{2}}\right)
\right).
\eeqn
The amplitudes $A_{(+,+;+,-)}$ and $A_{(+,-;+,+)}$ have a similar form,
the final factors being replaced by $C_1(Q_1) C_2(Q_2)$ and
$C_2(Q_1) C_1(Q_2)$ respectively.

\subsection{Numerical results}
In Fig.3 we show the results for the 
$\gam^{*}\gam^{*}\rightarrow\gam\gam$ differential cross-section 
obtained using equations (\ref{nonflip1}) and (\ref{flip1})
and compare with the results obtained using the approximations to
$I^{\nu}_{(+,+)}$ and $I^{\nu}_{(+,-)}$ given in (\ref{nonfliplim1}),
(\ref{fliplim1}),(\ref{nonfliplim2}) and (\ref{fliplim2}). All $\nu$ 
integrals are computed numerically.

At LEP2, $z$ values around 1 are typical and values of around 2 are more
typical of those attainable at a future linear collider.

\section{$\gam^{*}P\rightarrow\gam X$}
We may use the results (\ref{nonflip1}) and (\ref{flip1}) 
to compute $\gam^{*}\rightarrow\gam$ scattering off a hadronic state. 
We consider $\gam^{*}P\rightarrow\gam X$ scattering at the energies 
typical of those at the HERA collider.  

The impact factor for the coupling to the quarks and gluons in the proton
is well known \cite{MT,BFRLLW} and yields
\be\label{line}
I^{\nu}_{q}=-2(2\pi)^{5}\alpha_{s}\frac{\Gam(1/2-i\nu)}{\Gam(1/2+i\nu)}
\left(\frac{2}{q}\right)^{1-2i\nu}
\ee
for the coupling to quarks. The colour factor for coupling to quarks is 
$$G = \frac{N_{c}^{2}-1}{4 N_c} = 2/3,$$ relative to which the coupling to 
gluons is enhanced by a factor $C_A/C_F = 9/4$. 
Putting the colour factor, along with 
(\ref{nonflip1}),(\ref{flip1}),(\ref{line}), into (\ref{ampii}) yields
\beqn\label{ampGPnon}
A_{(+,+)}(s,t,Q^{2})&=&i\ \alpha \alpha_s^2 \sum_{q} e_q^2 \
\frac{\pi}{6}\ \frac{s}{|t|}\ \int d\nu\ \frac{\nu^2}{(1/4+\nu^2)^2}\
\frac{1}{1+\nu^2} \frac{\tanh(\pi \nu)}{\pi \nu}\ e^{z\chi(\nu)}
\nonumber \\ & & \hspace*{-3.2cm} 
\int_{\gam-1-i\infty}^{\gam-1+i\infty}\frac{dz}{2 \pi i}
\left(\frac{Q^2}{|t|}\right)^{-z/2} 
\frac{\Gam(1
/2-i\nu-z/2)\;\Gam(1/2+i\nu-z/2)}{\Gam(1/2+i\nu) \
\Gam(1/2-i\nu)}\ \Gam(z/2)\ \Gam(z/2+1)[z^{2}+11+12\nu^2] \nonumber \\
\eeqn
and
\beqn\label{ampGPflip}
A_{(+,-)}(s,t,Q^{2})&=&i\ \alpha \alpha_s^2 \sum_{q} e_q^2 \
\frac{2 \pi}{3}\ \frac{s}{|t|} \int d\nu\ \frac{\nu^2}{(1/4+\nu^2)^2}\
\frac{1}{1+\nu^2} \frac{\tanh(\pi \nu)}{\pi \nu}\
e^{z\chi(\nu)} \nonumber \\ & & \hspace*{-2.5cm}
\int_{\gam-1-i\infty}^{\gam-1+i\infty}\frac{dz}{2 \pi i}
\left(\frac{Q^2}{|t|}\right)^{-z/2} 
\frac{\Gam(3/2-i\nu-z/2)
\ \Gam(3/2+i\nu-z/2)}{\Gam(1/2+i\nu)
\ \Gam(1/2-i\nu)}\ \Gam(z/2)\ \Gam(z/2+1).
\eeqn

\subsection{The low $Q^2$ limit}
In the limit $-t \gg Q^2$ the saddle point integration about $\nu=0$ 
yields
\be  \label{gpQ2lltnon}
A_{(+,+)}(s,t,Q^{2}){\Big |}_{Q^{2}\ll |t|} = \frac{8i}{3}\
\alpha \alpha^{2}_{s} \sum_{q} e_q^2 \
\frac{s}{|t|}\ e^{4z\ln{2}}\left(\frac{\pi}{14z\zeta(3)}\right)^{3/2}
C_1(Q)
\ee
and
\be\label{gpQ2lltflip}
A_{(+,-)}(s,t,Q^{2}){\Big |}_{Q^{2}\ll |t|}=\frac{8i}{3}\
\alpha\alpha^{2}_{
s} \sum_{q} e_q^2 \
\frac{s}{|t|}\ e^{4z\ln{2}}\left(\frac{\pi}{14z\zeta(3)}\right)^{3/2}
C_2(Q).
\ee

\subsection{The high $Q^2$ limit}
The relevant saddle point is now at $\nu\approx\pm i/2$ and so
\be\label{gpQ2ggtnon}
A_{(+,+)}(s,t,Q^{2}){\Big |}_{Q^{2}\gg |t|}=\frac{8 i}{3\sqrt\pi}\
\alpha\alpha_{s}^{2} \sum_{q} e_q^2 \ \frac{s}{Q^2}\frac{
\ln^{3/4}(Q^{2}/|t|)}{z^{5/4}}\exp\left(2\sqrt{z \ln(Q^{2}/|t|)}\right)
\ee
and
\be\label{gpQ2ggtflip}
A_{(+,-)}(s,t,Q^{2}){\Big |}_{Q^{2}\gg |t|
}=\frac{16i}{9\sqrt\pi}\
\alpha\alpha_{s}^{2} \sum_{q} e_q^2 \
\frac{s |t|}{Q^4}\frac{
\ln^{3/4}(Q^{2}/|t|)}{z^{5/4}}\exp\left(2\sqrt{z \ln(Q^{2}/|t|)}\right)
\ee
The dominance of the non-flip amplitude over the flip amplitude is clear:
\be
\frac{A_{(+,+)}}{A_{(+,-)}}= \frac{3 Q^2}{2 |t|}.
\ee
\subsection{Numerical Results}
In Fig.4 we show the results for proton-photon scattering in 
by performing the $\nu$ and $z$ integrals exactly. We compare with the
results obtained using the analytic approximations to the amplitude 
derived from (\ref{fliplim1}),(\ref{nonfliplim1}),(\ref{fliplim2}),
(\ref{nonfliplim2}) again with the $\nu$ integral performed exactly.  

The proton-photon cross section is determined by convoluting the
photon-quark cross section with the parton density functions 
\cite{pdflib,MRS}:
\be
\frac{d\sigma(\gam P \rightarrow \gam X)}{dt}=
\int_{x_{{\rm min}}}^1 dx \left(\frac{81}{16} g(x,t)+\sum_f
(q(x,t)+\overline{q}(x,t))\right)
\frac{d\sigma(\gam q \rightarrow \gam q)}{dt}.
\ee
The lower limit of the $x$ integral is determined by the on-shell condition 
for the struck quark in the proton, i.e.
\be
x_{{\rm min}}=\frac{1}{(\frac{M_X^2}{|t|}+1)},
\ee
where $M_X$ is the maximum invariant mass of the proton dissociation products.
We choose $M_X=10$ GeV, which is typical of the cuts imposed by the HERA
experiments. We note that in the limit of $|t|\gg Q^2$ the cross section 
becomes insensitive to values of $Q^2$, as anticipated in the analytic 
expressions (\ref{gpQ2lltnon}) and (\ref{gpQ2lltflip}). The approximate
expressions are shown to be good over a large range in $Q^2/|t|$. 
We note that these cross-sections are large enough to make observation at
HERA feasible.

\section{Acknowledgements}
We wish to thank Mark W\"usthoff for many long and detailed discussions. 
NGE was funded for this work from a PPARC studentship. 
This work was supported in part by the EU Fourth Programme `Training and 
Mobility of Researchers', Network `Quantum Chromodynamics and the Deep 
Structure of Elementary Particles', contract FMRX-CT98-0194 (DG 12-MIHT).

\begin{center}
\begin{tabular}{|c|c|}\hline
\epsfig{file=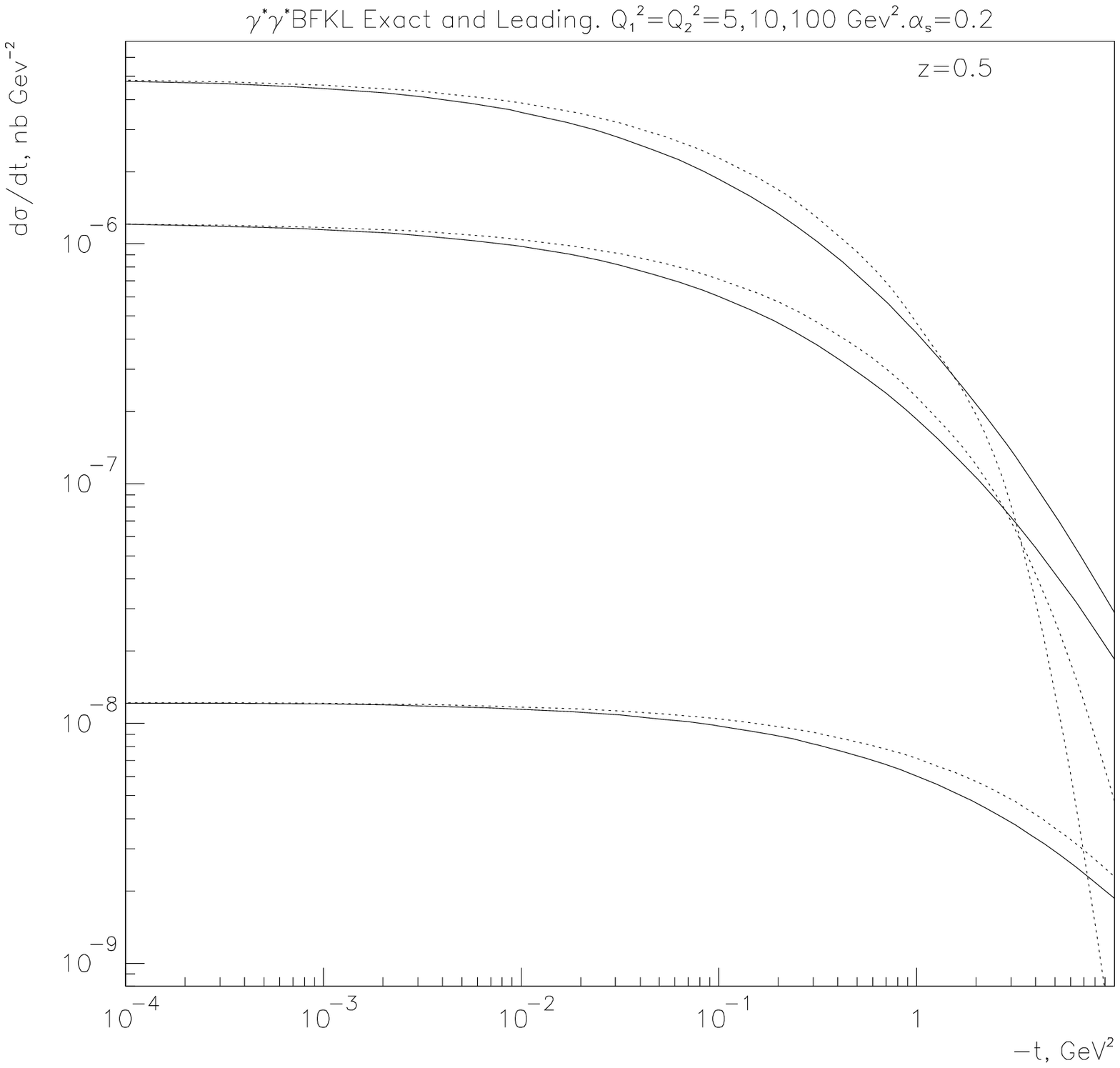,width=7.2cm}&
\epsfig{file=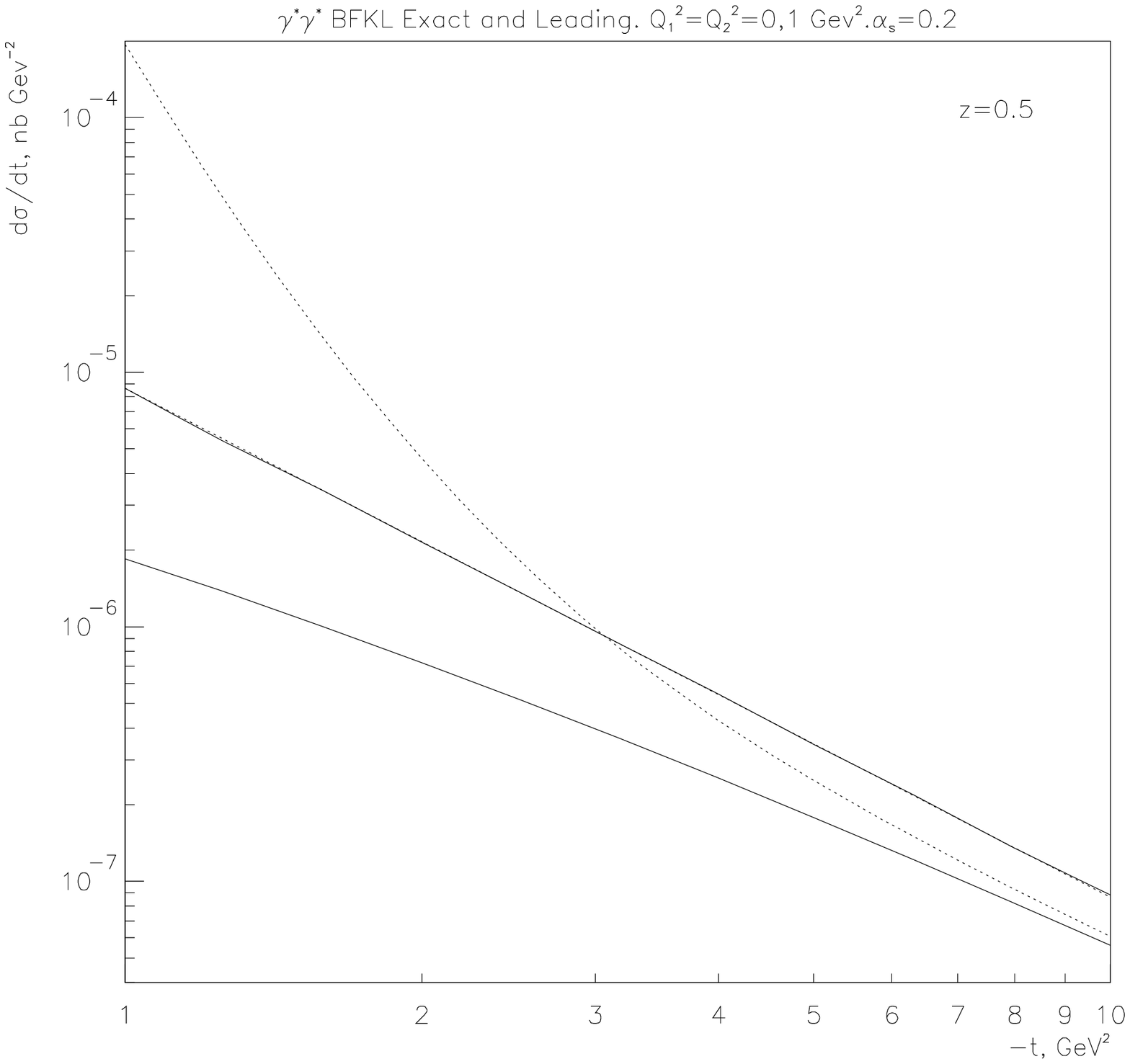,width=7.2cm}\\ \hline
\epsfig{file=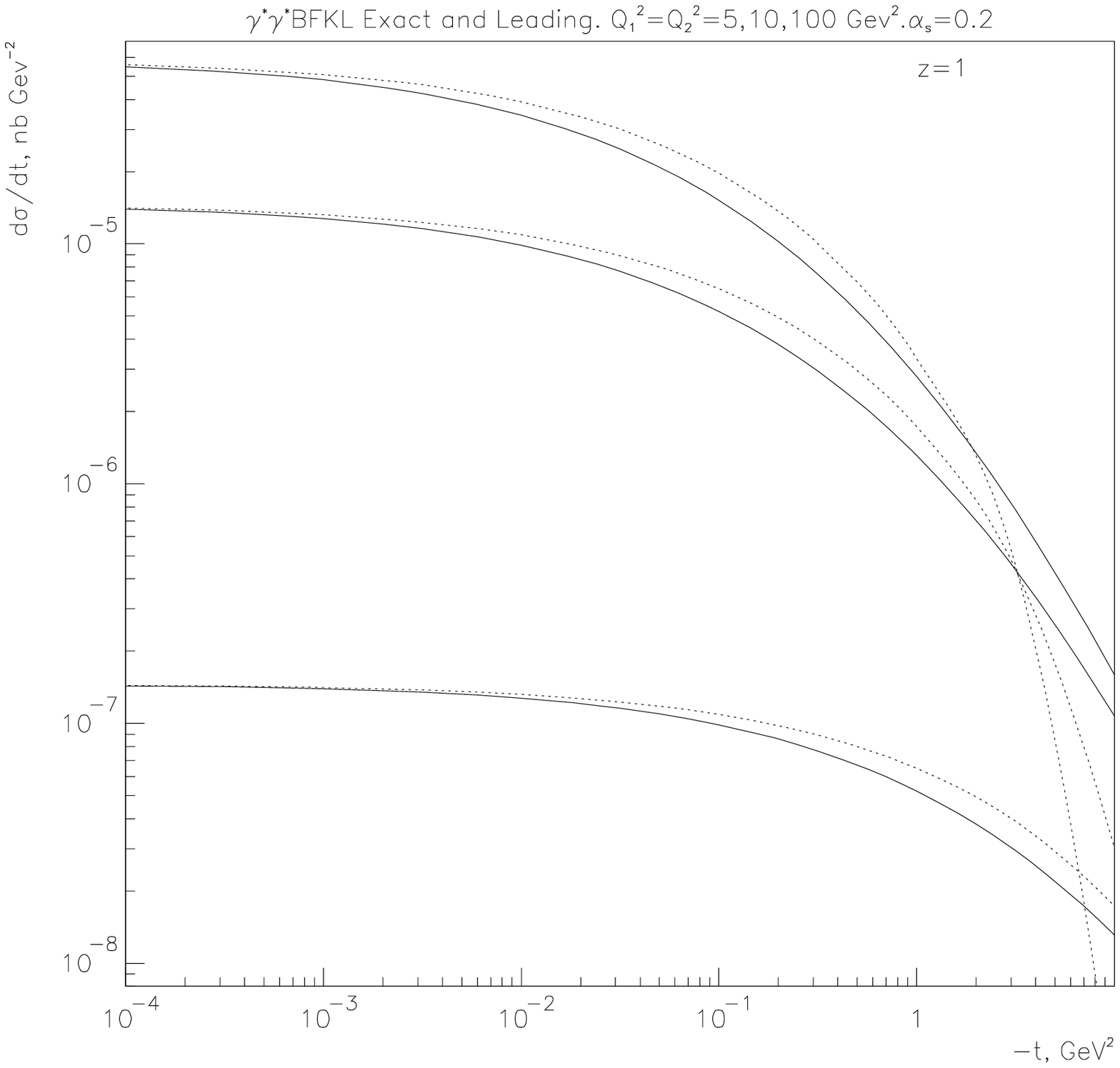,width=7.2cm}&
\epsfig{file=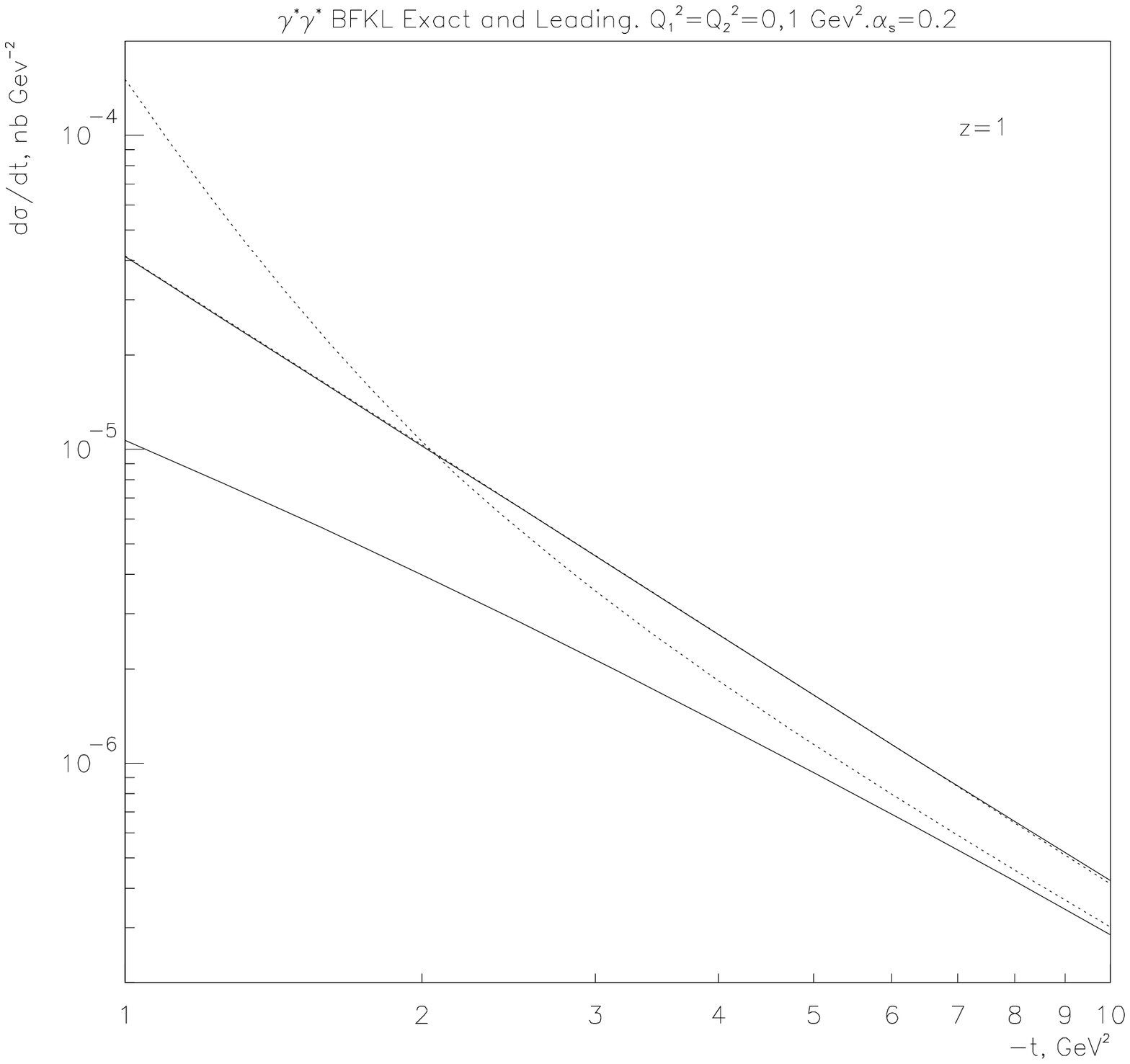,width=7.2cm}\\ \hline
\epsfig{file=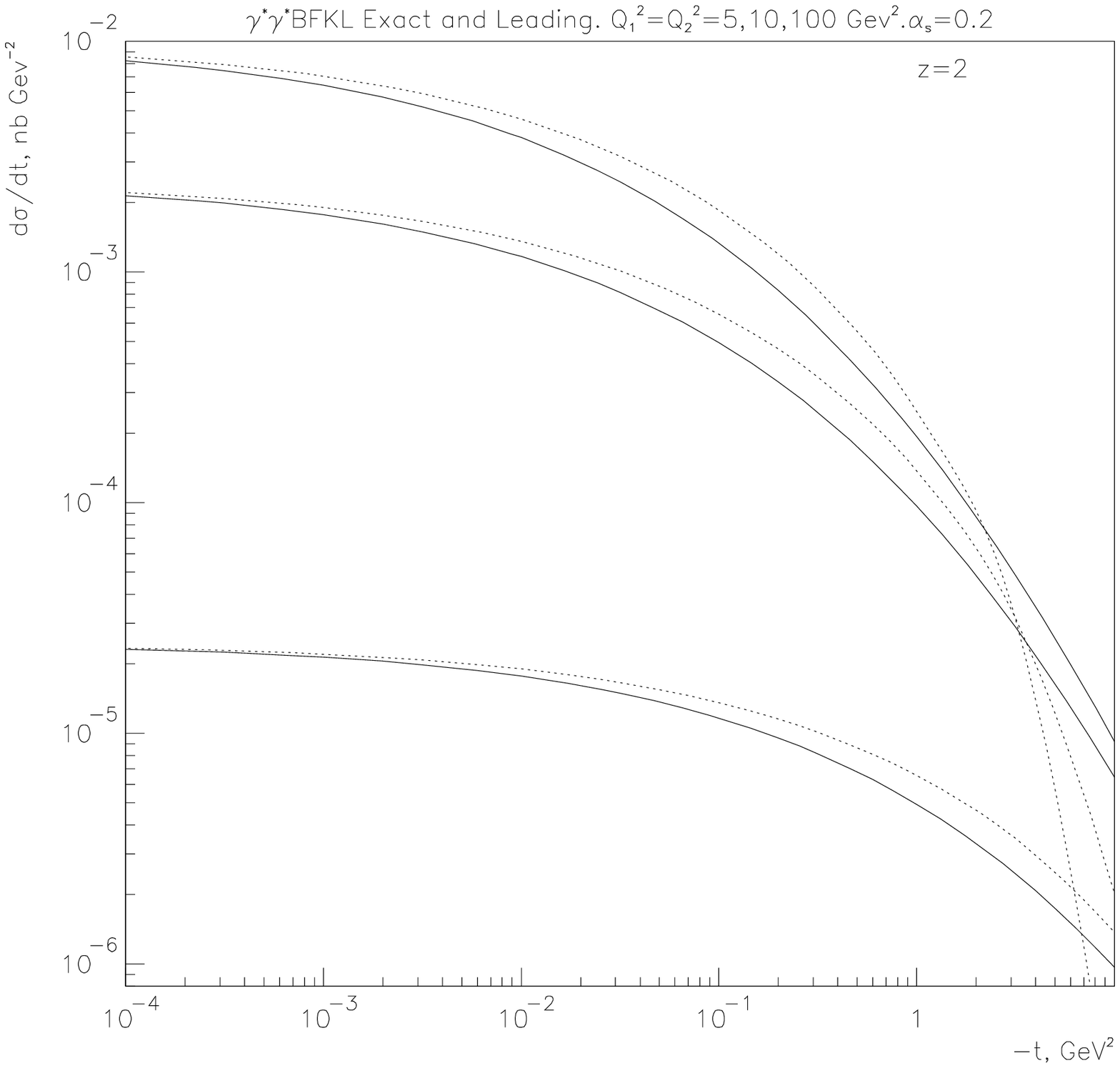,width=7.2cm}&
\epsfig{file=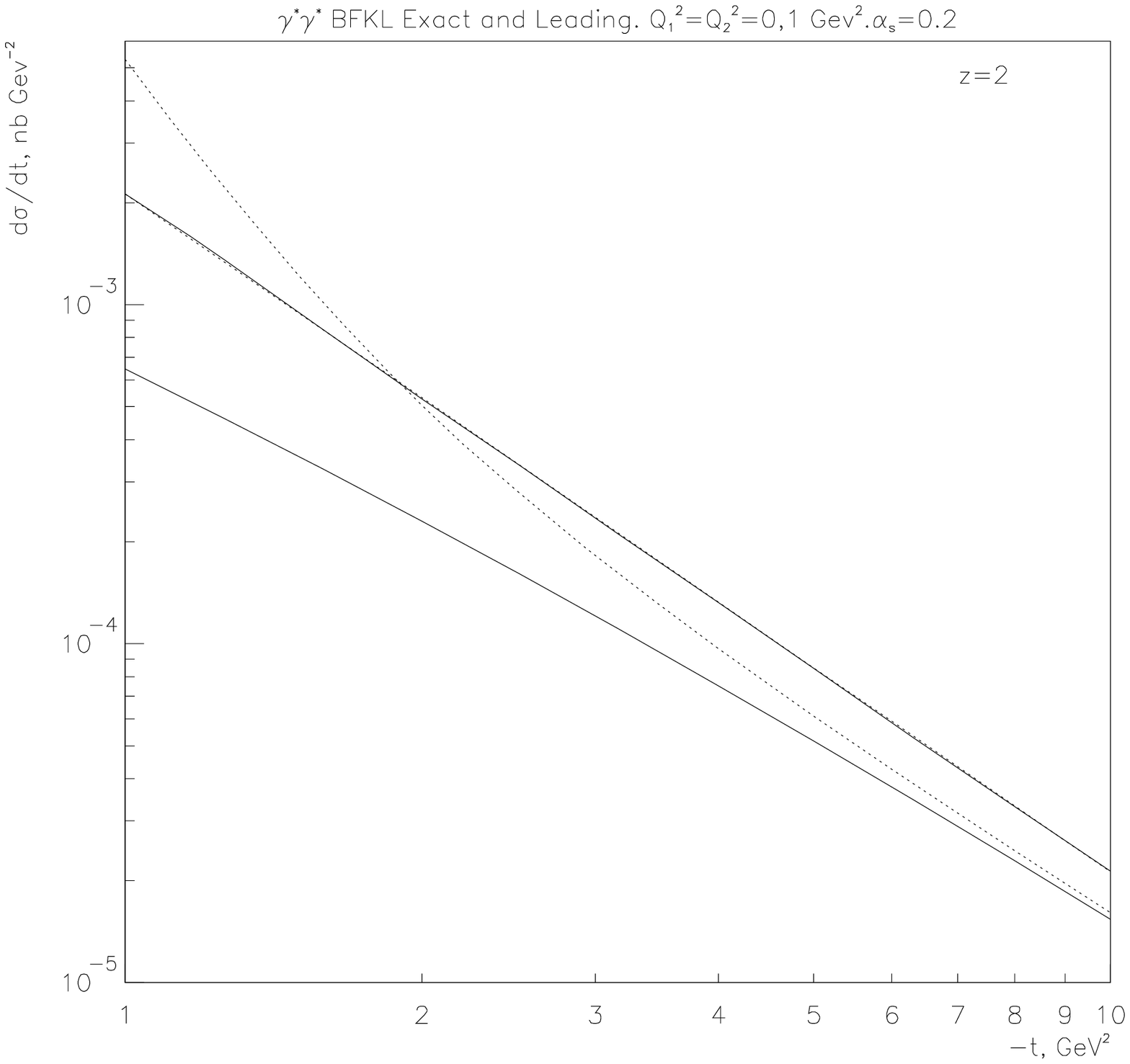,width=7.2cm}\label{GGfig}\\ \hline
\end{tabular}
\end{center}
\begin{center}
Figure 3. $\gam^{*}\gam^{*}\rightarrow\gam\gam$ cross section comparing the exact BFKL result with
the leading approximation. The solid line gives the exact result and the dotted line the result obtained by only keeping 
the leading terms as described in the text.
\end{center}
\begin{center}
\begin{tabular}{|c|c|}\hline
\epsfig{file=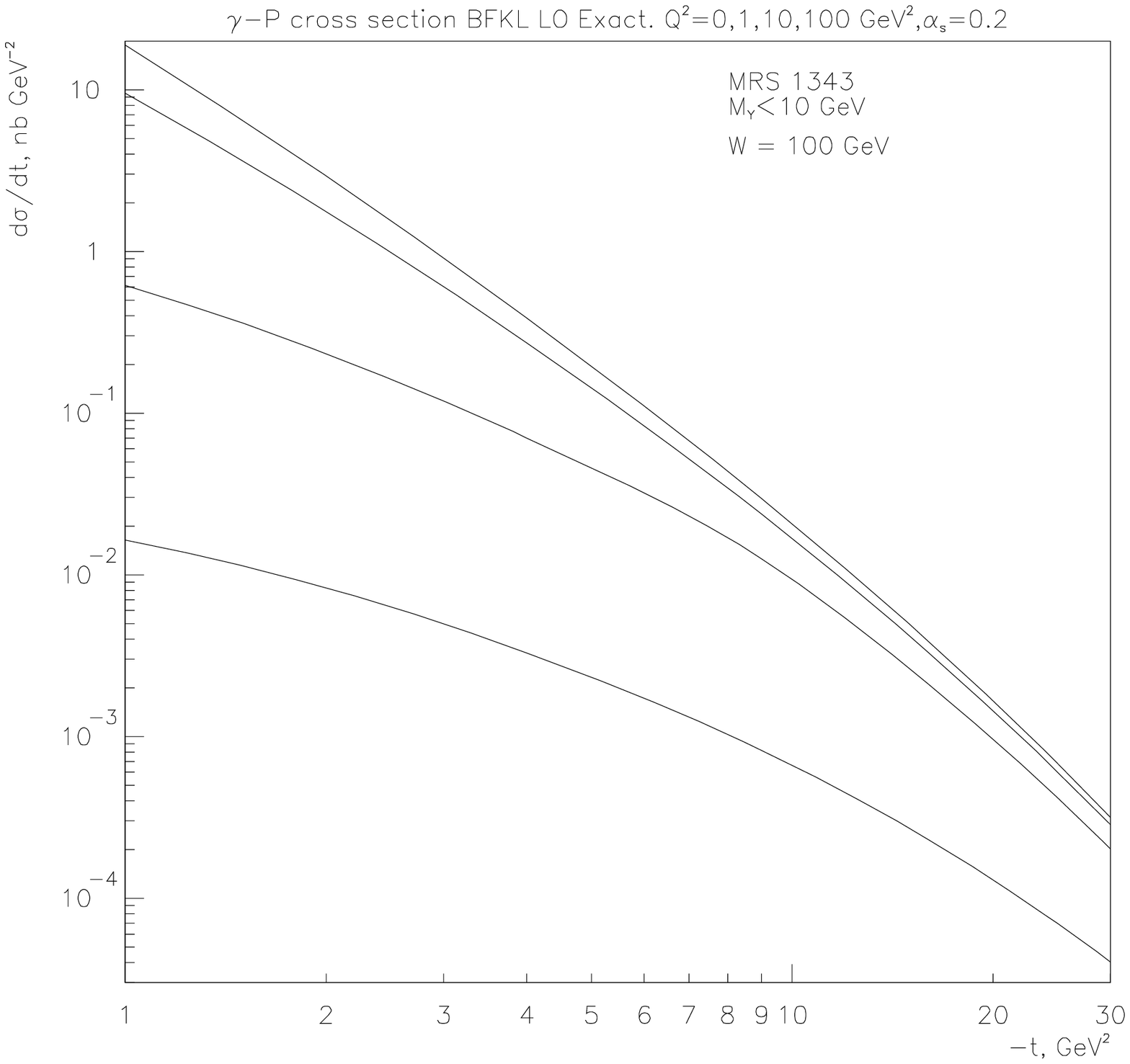,width=7.2cm}&
\epsfig{file=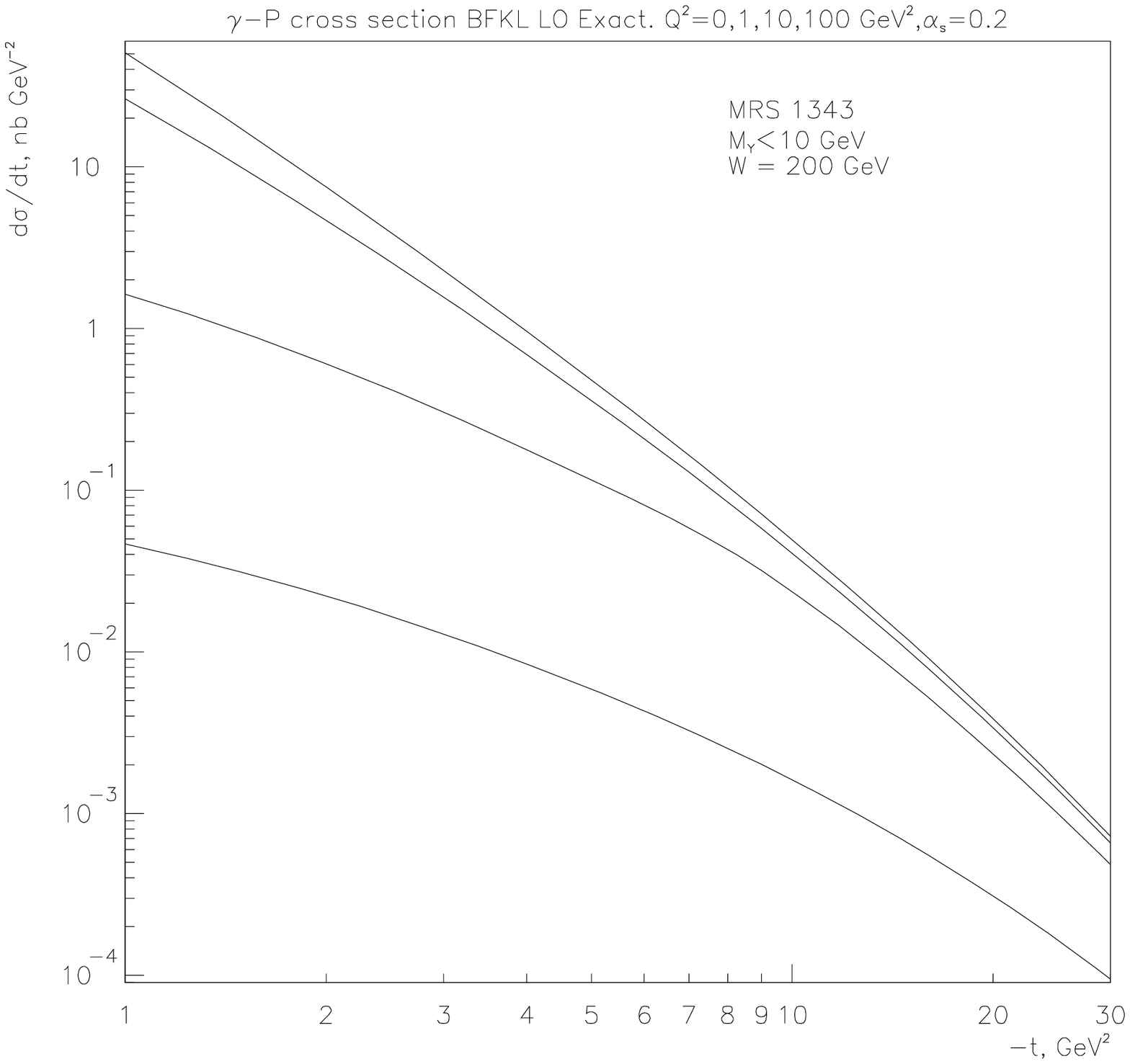,width=7.2cm}\\ \hline
\epsfig{file=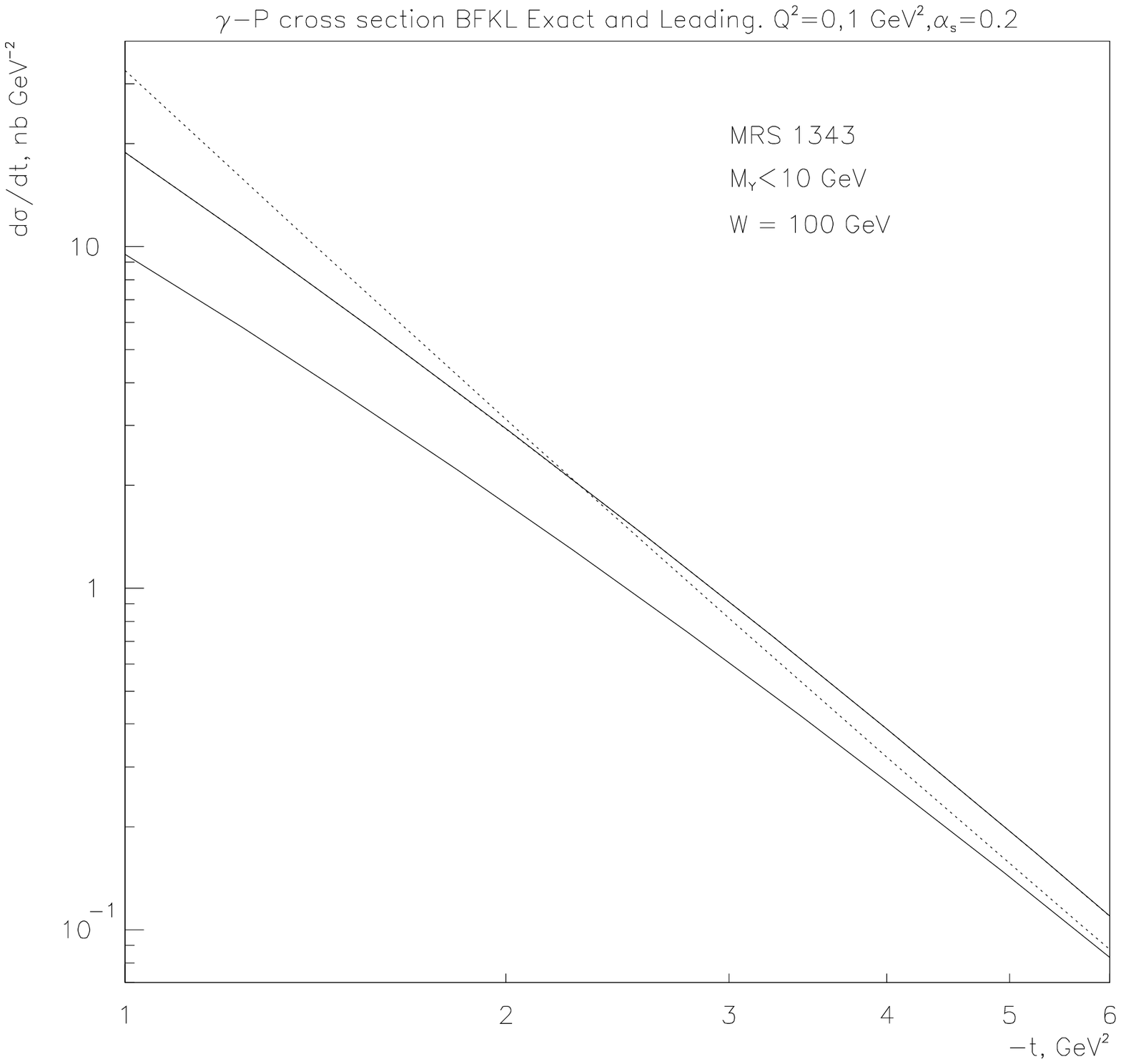,width=7.2cm}&
\epsfig{file=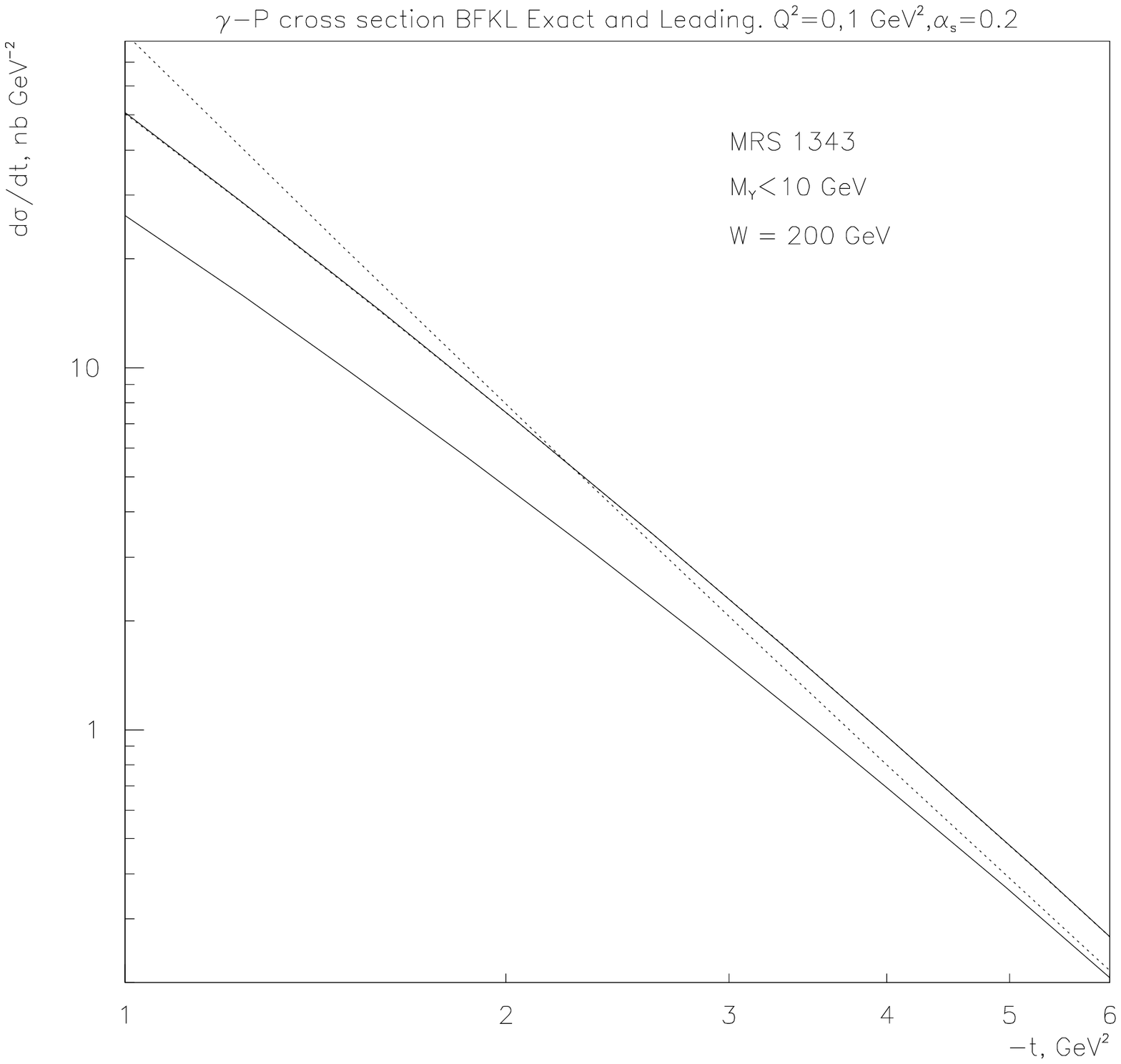,width=7.2cm}\\ \hline
\epsfig{file=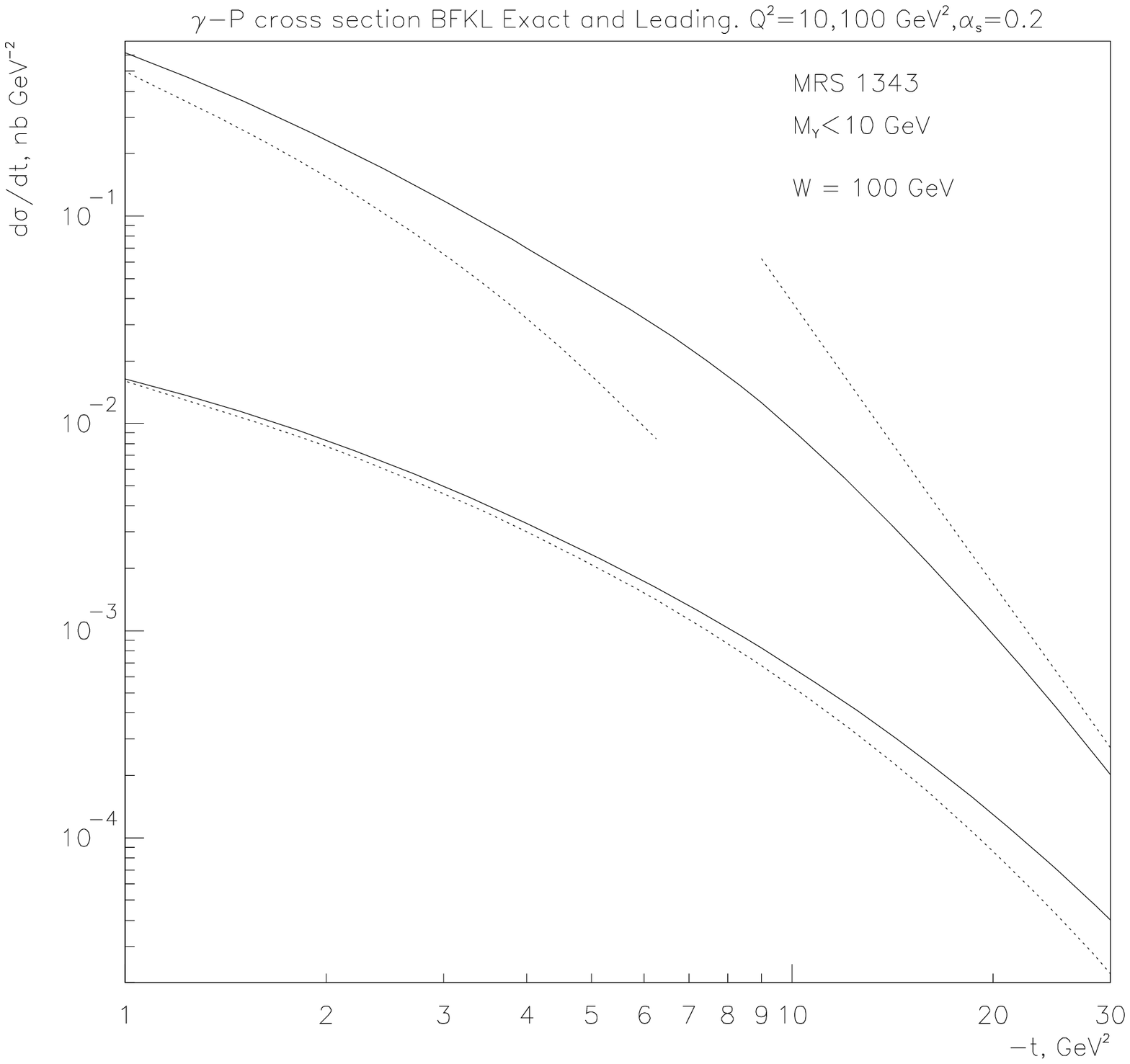,width=7.2cm}&
\epsfig{file=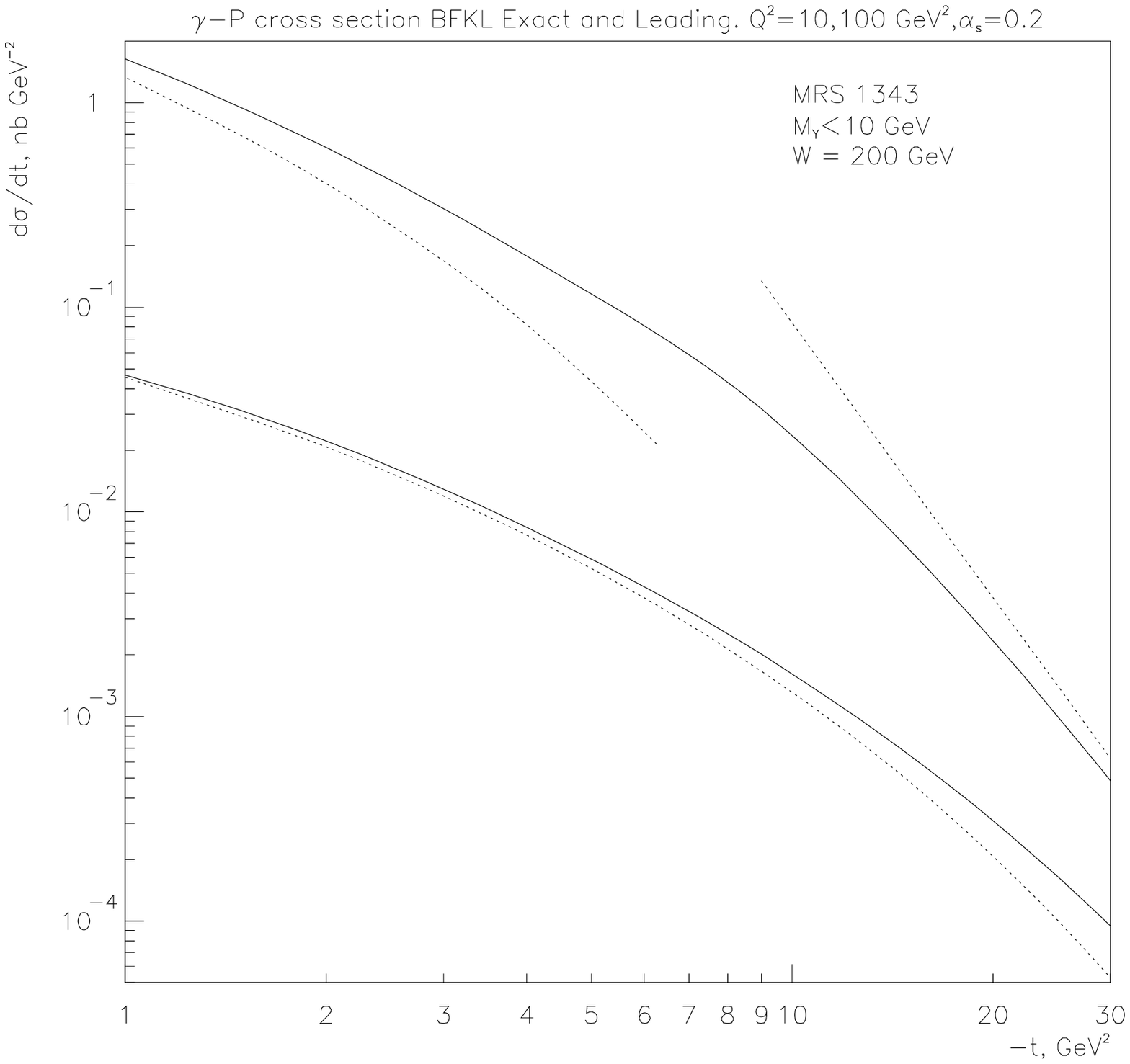,width=7.2 cm}\label{GPfig}\\ \hline
\end{tabular}
\end{center}
\begin{center}
Figure 4. $\gam^{*}P\rightarrow\gam X$ cross section comparing the exact BFKL result with
the leading approximation. The solid line gives the exact result and the dotted line the result obtained by only keeping 
the leading terms as described in the text.
\end{center}


\begin{thebibliography}{99}
\bibitem{FS} J.R.~Forshaw and P.J.~Sutton, Euro.~Phys.~J {\bf C14} (1998) 285.
\bibitem{MT} A.H.~Mueller and W-K.~Tang, Phys.~Lett. {\bf B284} (1992) 123. 
\bibitem{CF} B.E.~Cox and J.R.~Forshaw, Phys.~Lett. {\bf B434} (1998) 133. 
\bibitem{jets} ZEUS Collaboration: M.Derrick et al, Phys.~Lett {\bf B369} (1996) 55; \\
H1 Collaboration: ``Rapidity Gaps Between Jets in Photoproduction at HERA'' contribution to The International Europhysics Conference on High Energy Physics, Jerulsalem, Israel, August 1997;  \\
H1 Collaboration: ``Double Diffraction Dissociation at large $|t|$ in Photoproduction at HERA'' contribution to ICHEP98, Vancouver, Canada, July 1998;\\
CDF Collaboration: F.Abe. et al, Phys.~Rev.~Lett. {\bf 74} (1995) 855; \\
D0 Collaboration: S.~Abachi et al, Phys.~Rev.~Lett. {\bf 76} (1996) 734.
\bibitem{Ginzburg} I.F.~Ginzburg, S.L.~Panfil, V.G.~Serbo, Nucl.~Phys {\bf B284} (1987) 685.
\bibitem{GinzburgIvanov}I.F.~Ginzburg and D.Yu.~Ivanov, Phys.~Rev. {\bf D54} (1996) 5523.
\bibitem{FRIvanovBFHL} J.R.~Forshaw and M.G.~Ryskin, Zeit.~Phys. {\bf C68} (1995) 137;\\
J.~Bartels, J.R.~Forshaw, H.~Lotter, M.~W\"usthoff, Phys.~Lett. {\bf B375} (1996) 301;\\
D.Yu.~Ivanov, Phys.~Rev. {\bf D53} (1996) 3564;\\
D.Yu.~Ivanov and R.~Kirschner, Phys.~Rev. {\bf D58} (1998) 4026. 
\bibitem{hight} H1 Collaboration: ``Production of $J/\Psi$ Mesons with large $|t|$ at HERA'' contribution to The International 
Europhysics Conference, Jerusalem, Israel, August 1997;  \\
ZEUS Collaboration: ``Study of Vector Meson Production at Large $|t|$ at HERA''
contribution to The International Europhysics Conference, Jerusalem, Israel, August 1997;\\
ZEUS Collaboration: ``Study of Vector Meson Production at Large $|t|$ at HERA and Determination of the Pomeron Trajectory'' contribution to ICHEP98, Vancouver, Canada, July 1998.
\bibitem{MWDI} D.Yu.~Ivanov  and M.~W\"usthoff,  preprint hep-ph/9808455. 
\bibitem{BRLBrodsky} J.~Bartels, A.~De~Roeck, H.~Lotter, Phys.~Lett. {\bf B389} (1996) 742; \\
S.J.~Brodsky, F.~Hautmann, D.E.~Soper, Phys.~Rev. {\bf D56} (1997) 6957; \\
S.J.~Brodsky, F.~Hautmann, D.E.~Soper, Phys.~Rev.~Lett. {\bf 78} (1997) 803; Erratum-ibid. {\bf 79} (1997) 3544.
\bibitem{bfkl} Ya.Ya.~Balitsky and L.N.~Lipatov, Sov. J. Nucl.~Phys. {\bf 28} (1978) 822; \\
E.A.~Kuraev, L.N.~Lipatov, V.S. ~Fadin, Phys.~Lett. {\bf B60} (1975) 5; Sov.~Phys.~JETP {\bf 71} (1976) 840; Sov. Phys. JETP {\bf 72} (1977) 377.
\bibitem{BFKLsolve} L.N.~Lipatov, Sov.~Phys.~JETP {\bf 63} (1986) 904.
\bibitem{LipQED} L.N.~Lipatov and G.V.~Frolov, Sov. J.~Nucl.~Phys. {\bf 13.3} (1971) 333; \\ 
H.~Cheng and T.T.~Wu, Phys.~Rev. {\bf D1.123} (1970) 3414.
\bibitem{JRFDR} J.R.~Forshaw and D.A.~Ross, `QCD and the Pomeron', Cambridge University Press (1997).
\bibitem{FORM} J.A.M.~Vermaseren, `Symbolic Manipulation with FORM, Tutorial and Reference Manual', CAN\footnote{e-mail:form@can.nl}  (1991). 
\bibitem{BFRLLW} J.~Bartels, J.R.~Forshaw, H.~Lotter, L.N.~Lipatov, M.G.~Ryskin, M.~W\"usthoff, Phys.~Lett. {\bf B348} (1995) 589.
\bibitem{pdflib}  H.~Plothow-Besch, Comp.~Phys.~Comm. {\bf 75} (1993) 396;\\
H.~Plothow-Besch, Int.~J.~Mod.~Phys. {\bf A10} (1995) 2901.
\bibitem{MRS} A.D.~Martin, R.G.~Roberts, W.J.~Stirling, Phys.~Rev. {\bf D51} (1995) 4756.
\end{thebibliography}
\end{document}